\documentclass[a4paper,11pt]{article}
\pdfoutput=1 % if your are submitting a pdflatex (i.e. if you have
\usepackage{jheppub} % for details on the use of the package, please
\usepackage[T1]{fontenc} % if needed

\usepackage{bm}
\usepackage{mathrsfs} % \mathscr{L}
\usepackage{slashed} % Feynman slash
\usepackage{stackrel}
\usepackage{enumerate}
\usepackage{listings} % code snippets

\newcommand{\MSbar}{$\overline{\text{MS}}$}

\graphicspath{
    {./}
    {./figs/}
    {./figs/equilibrium/}
    {./figs/gravitational_waves/}
}

\title{Perturbative gravitational wave predictions for the real-scalar extended Standard Model}
\author{Oliver Gould}
\emailAdd{oliver.gould@nottingham.ac.uk}
\author{and Paul M. Saffin}
\emailAdd{paul.saffin@nottingham.ac.uk}
\affiliation{School of Physics and Astronomy, University of Nottingham, Nottingham NG7 2RD, United Kingdom}
\date{\today}

\abstract{
    We perform a state-of-the-art study of the cosmological phase transitions of the real-scalar extended Standard Model. We carry out a broad scan of the parameter space of this model at next-to-next-to-leading order in powers of couplings. We use effective field theory to account for the necessary higher-order resummations, and to construct consistent real and gauge-invariant gravitational wave predictions. Our results provide a comprehensive account of the convergence of perturbative predictions for the gravitational wave signals in this model. For the majority of the parameter points in our study, we observe apparent convergence. While leading and next-to-leading order predictions of the gravitational wave amplitude typically suffer from relative errors between $10$ and $10^4$, at next-to-next-to-leading order the typical relative errors are reduced to between $0.5$ and $50$. Nevertheless, for those parameter points predicting the largest signals, potentially observable by future gravitational wave observatories, the validity of the perturbative expansion is in doubt.
}

\begin{document}
\maketitle
\flushbottom

%%%%%%%%%%%%% Introduction %%%%%%%%%%%%%
\section{Introduction} \label{sec:introduction}

Gravitational wave observatories offer the exciting prospect to be able to probe the very early universe, through sensitivity to the gravitational waves produced by a cosmological first-order phase transition. However, recent works have cast doubt on the reliability of current gravitational wave predictions within perturbation theory, finding uncertainties in the expected amplitude of several orders of magnitude~\cite{Croon:2020cgk}. The scalar singlet extension of the Standard Model has been at the centre of this debate~\cite{Carena:2019une, Gould:2021oba, Athron:2022jyi, Lewicki:2024xan, Ramsey-Musolf:2024ykk}. In the $\mathrm{Z}_2$ symmetric version of this model, some groups have found phase transitions to be observable by planned detectors, and others not, with the differences apparently due to loop corrections.

For the Standard Model (SM), it has long been clear that perturbation theory fails to describe the electroweak phase transition even qualitatively. A loop expansion of the Higgs effective potential predicts a second-order transition at one-loop (unresummed), but a first-order transition at higher loops, all in contradiction with lattice simulations, which find a crossover~\cite{Kajantie:1995kf, Kajantie:1996mn}. This is despite the electroweak sector being weakly coupled at zero temperature. At high temperatures near the electroweak phase transition, the long-wavelength modes of the Higgs and $W$ and $Z$ bosons become strongly coupled~\cite{Linde:1980ts}, and it is these modes that determine the nature of the phase transition~\cite{Kajantie:1995dw}.

Nevertheless, a perturbative expansion can be constructed for an electroweak-like phase transition. One finds that the effective expansion parameter is $\sim 20\lambda/g^2$, where $\lambda$ is the Higgs self-coupling, $g^2$ is the weak gauge coupling and the approximate value 20 follows from the coefficients of the expansion~\cite{Ekstedt:2022zro, Ekstedt:2024etx}. This expansion parameter is greater than one for Standard Model couplings, indicating that the perturbative expansion does not converge. For small enough values of the Higgs self-coupling such that the expansion is convergent, the transition is strongly first order. In this case, lattice and perturbative predictions are in good agreement~\cite{Gould:2022ran, Ekstedt:2022zro, Ekstedt:2024etx}. Thus, one may ask: can perturbation theory describe the gravitational waves produced by strongly first-order phase transitions in extensions of the Standard Model? The present paper aims to address this question in the context of a scan of the parameter space of a concrete model, building on and complementing refs.~\cite{Gould:2021oba, Gould:2023jbz, Lewicki:2024xan, Ramsey-Musolf:2024ykk}.

The Lagrangian for the real singlet scalar extension of the Standard Model (xSM) is
\begin{align} \label{eq:lagrangian_xsm}
    \mathscr{L}               & = \mathscr{L}_\text{SM} + \mathscr{L}_s + \mathscr{L}_\text{portal} ,          \\
    \mathscr{L}_s=            & - \frac{1}{2}\partial_\mu s \partial^\mu s
    - b_1 s
    - \frac{1}{2}m_s^2 s^2
    - \frac{1}{3}b_3 s^3
    - \frac{1}{4}b_4 s^4,                                                                                      \\
    \mathscr{L}_\text{portal} & = - \frac{1}{2}a_1 s \phi^\dagger \phi - \frac{1}{2}a_2 s^2 \phi^\dagger \phi,
\end{align}
where $\mathscr{L}_\text{SM}$ is the Standard Model Lagrangian, $\phi$ is the Higgs field and $s$ is the new scalar. This is the most general renormalisable Lagrangian for this field content. Our notation largely follows ref.~\cite{Niemi:2021qvp}.

The xSM is an ideal playground for the study of cosmological phase transitions, as it can yield a wide variety of different cosmological phase histories including strong first-order phase transitions. It is also motivated as a minimal extension of the Standard Model that provides a dark matter candidate~\cite{Silveira:1985rk, GAMBIT:2017gge}, and may provide for baryogenesis together with higher-dimension CP-violating operators~\cite{Cline:2012hg, Cline:2021iff, Harigaya:2022ptp, Ellis:2022lft}. If the new scalar has a mass much below the electroweak scale, it can also be relevant in astrophysical and cosmological contexts~\cite{Burrage:2018dvt, Brax:2021rwk}.

We study the cosmological thermal history of this model, focusing on temperatures in the vicinity of the electroweak phase transition. In the Standard Model the pseudocritical temperature of the crossover transition is approximately $160~\mathrm{GeV}$~\cite{DOnofrio:2015gop}. Our aim is to study the reliability and convergence of perturbation theory to describe possible phase transitions in this model and their gravitational wave signals.

Ref.~\cite{Carena:2019une} studied phase transitions in the $Z_2$ symmetric xSM ($b_1=b_3=a_1=0$), and demonstrated that at leading, one-loop order, the model predicts observable gravitational wave signals. However, given the importance of the one-loop terms, their effects should be resummed. At leading order, doing so corresponds to including the infinite set of daisy diagrams~\cite{Arnold:1992rz}. Carrying out this resummation, ref.~\cite{Carena:2019une} found that the predicted gravitational wave signals were reduced in amplitude by 10 orders of magnitude, rendering them unobservable by planned detectors.

As thermal effects, and consequently phase transitions, arise first at one-loop order, a study of convergence requires reaching at least (resummed) two-loop order. The first two-loop study of phase transitions and gravitational wave signals in the xSM was carried out in ref.~\cite{Gould:2021oba}, building on refs.~\cite{Niemi:2021qvp, Schicho:2021gca}. The predicted gravitational wave amplitudes were found to be observable, but at leading order suffered from up to 11 orders of magnitude uncertainty. At two-loop order, these uncertainties were significantly reduced, to between 1 and 4 orders of magnitude, and the overall amplitude of the signal was increased. However, this study was limited to just two parameter points, so it was unclear to what extent the conclusions were special to the parameter points chosen.

A number of recent works have extended and consolidated this line of study, by carrying out two-loop order analyses of phase transitions in the xSM for scans of the model parameter space. Refs.~\cite{Niemi:2024vzw, Niemi:2024axp} have done so for the equilibrium phase transition properties, and refs.~\cite{Lewicki:2024xan, Ramsey-Musolf:2024ykk} for the gravitational wave signals. Ref.~\cite{Niemi:2024axp} also carried out lattice Monte-Carlo simulations, to definitively test the validity of perturbative predictions. Our work extends and complements these studies.

In this paper, we analyse the convergence of perturbation theory for gravitational wave signals of phase transitions in the xSM. We carry out our analysis at a whole model level, by performing a comprehensive scan of the generic (non-$Z_2$-symmetric) parameter space at three successive perturbative orders. At each order, we use renormalisation scale dependence as a proxy for the theoretical uncertainty of the prediction, and we quantify the reduction of uncertainty at higher orders. We analyse how the distribution of predictions, as well as the distribution of uncertainties, changes between orders in a given parameter scan.

%%%%%%%%%%%%% Methods %%%%%%%%%%%%%
\section{Methods} \label{sec:methods}

In short, we perform coupling expansions for phase transition properties of interest, assuming high temperatures. The study of thermal phase transitions requires resummations of the loop expansion to construct the coupling expansion, for which we utilise power counting and effective field theory (EFT). In this way, the calculation of phase transition properties is split into two parts: first we construct the EFT in Section~\ref{sec:eft}, then we make calculations within it in Section~\ref{sec:analysing_phase_transitions}. For the latter, we adopt some perturbative methods that have proven effective in benchmark comparisons to lattice Monte-Carlo simulations. The particulars are outlined in what follows and references therein.

%%%%%%%%%%% constructing the EFT %%%%%%%%%%%
\subsection{Constructing the effective field theory} \label{sec:eft}

In weakly coupled theories, phase transitions naturally take place at high temperatures $T$, where $\pi T$ is large compared to relevant mass scales, $m$. From the perspective of the free energy density, this is because thermal corrections are suppressed by small couplings $g^2/(4\pi)^2 \ll 1$ relative to tree-level vacuum terms, yet thermal corrections must become sizeable at a thermal phase transition. At higher temperatures, the size of thermal corrections are enhanced, and a hierarchy of scales $\pi T \gg m$ allows to overcome the relative suppression by small couplings.

Yet this hierarchy of scales leads to a breakdown of the vanilla loop expansion and requires resummation. While many approaches to this problem have been proposed in the literature, EFT has proved the most practical. It has been extended to the highest orders~\cite{Gould:2023jbz}, and can yield order-by-order gauge invariant predictions~\cite{Croon:2020cgk}. Further, perturbative expansions based on EFT appear to agree better with the results of lattice Monte-Carlo simulations than other approaches; see refs.~\cite{Blaizot:2003iq, Laine:2006cp} for QCD, and refs.~\cite{Ekstedt:2022zro, Gould:2023ovu} for electroweak studies.

%%%%%%%%%%% power counting %%%%%%%%%%%
\subsubsection{Power counting} \label{sec:power_counting}

Power counting allows to unambiguously identify the necessary resummations of the loop expansion required to construct a coupling expansion~\cite{Arnold:1992rz, Kajantie:1995dw, Braaten:1995cm}.
This follows from the uniqueness of the coefficients of an asymptotic expansion~\cite{Bender:1999box}. With EFT, the expansions at each energy scale can be carried out independently. For some recent discussion in this context see refs.~\cite{Lofgren:2023sep, Gould:2023ovu, Camargo-Molina:2024sde}.

We introduce a power counting parameter $x$ to define a weak-coupling limit $x\to 0_+$. Our power counting assumptions are then to take the larger dimensionless couplings in the Standard Model, and all new four-point couplings, to be of the same asymptotic order,
\begin{equation} \label{eq:power_counting_four_point}
    (4\pi)^2 x^2 \sim g^2 \sim g'^2 \sim g_\text{s}^2 \sim y_t^2 \sim \lambda \sim a_2 \sim b_4,
\end{equation}
where $\sim$ here means that the ratio of terms on each side of the relation goes to a finite, nonzero constant in this limit. The notation for the couplings is standard and follows ref.~\cite{Niemi:2021qvp}. The smaller Yukawa couplings of the Standard Model are neglected, the largest of which is $y_b^2/(4\pi)^2 \sim 8\times 10^{-4}$, of the bottom quark. The assumptions of equation~\eqref{eq:power_counting_four_point} are relatively standard, and equivalent assumptions were made in e.g.\ \cite{Kajantie:1995dw}.

For the new dimensionful three-point couplings, we take
\begin{equation}
    (4\pi)^2 x^2 \sim \frac{a_1^2}{m_s^2} \sim \frac{b_3^2}{m_s^2},
\end{equation}
which ensures that singlet loops involving three-point couplings are of the same order as those involving four-point couplings.

Together, these power counting assumptions ensure that the loop expansion and the coupling expansion agree in the absence of hierarchies of scale. For example, the loop expansion of the beta functions are then equivalent to an expansion in powers of $x^2$, with $O(1)$ coefficients. In practice, the convergence of the expansion is determined by the largest coupling, so that numerically this should be identified with $(4\pi)^2 x^2$.

Finally, we make the following power counting assumption about the scalar thermal effective masses,
\begin{equation} \label{eq:soft_assumption}
    (\pi T)^2 x^2 \sim m_{\phi,\text{eff}}^2 \sim m_{s,\text{eff}}^2,
\end{equation}
where $m_{\phi,\text{eff}}$ and $m_{s,\text{eff}}$ are the thermal effective masses of the Higgs and singlet scalar respectively. This assumption is equivalent to assuming the scalar fields live at the \textit{soft scale}, $x\pi T$, in the language of ref.~\cite{Gould:2023ovu}. Such an assumption arises naturally due to the form of the one-loop static thermal self-energies,
\begin{align}
    \Pi_\phi & \approx \frac{1}{(4\pi)^2}\left(3g^2 + g'^2 + 4y_t^2 + 8\lambda + \frac{2}{3}a_2\right)(\pi T)^2 , \\
    \Pi_s    & \approx \frac{1}{(4\pi)^2}\left(4 b_4 + \frac{8}{3}a_2\right)(\pi T)^2,
\end{align}
or more generally $\Pi \sim x^2 (\pi T)^2$. Thus, at sufficiently high temperatures, where the thermal self-energy is at least as large as the vacuum mass, the assumption of equation~\eqref{eq:soft_assumption} holds.

This set of power counting assumptions is relatively general and leads to a simple perturbative expansion. There are only two relevant energy scales to consider: the \textit{hard scale} $\pi T$ of the nonzero Matsubara modes, and the \textit{soft scale} $x\pi T$ of the zero Matsubara modes. In principle, there is also the nonperturbative \textit{ultrasoft scale} $x^2\pi T$, though this does not contribute to observables until higher orders than we consider~\cite{Linde:1980ts}. For the perturbative expansion of the free energy, the hard scale contributes integer powers of $x^2$, and the soft scale contributes both integer and half-integer powers of $x^2$. In total, the expansion of the free energy is then an expansion in integer powers of $x$.

Alternative power counting assumptions are possible, typically leading to more complicated perturbative expansions. In particular, the scalar fields can become parametrically lighter than their thermal self-energies, if there is an almost perfect cancellation between a negative tree-level mass term and the positive thermal self-energies. This happens when there is an apparent second-order phase transition, and leads to infrared divergences in the naive expansion. In this case, resolving the infrared divergences requires constructing a new EFT for the lower energy scale of the light scalar~\cite{Gould:2023ovu}. One natural possibility is then that $m_\text{eff}^2\sim x^3 (\pi T)^2$~\cite{Arnold:1992rz, Ekstedt:2020abj, Camargo-Molina:2021zgz, Ekstedt:2022zro, Gould:2023ovu, Ekstedt:2024etx, Kierkla:2023von}, leading to a dual perturbative expansion in powers of $x$ and $x^{3/2}$. For a broader discussion of the spectrum of possible power-counting assumptions see refs.~\cite{Lofgren:2023sep, Gould:2023ovu, Camargo-Molina:2024sde}. 
In this paper, we do not explore such alternative power counting assumptions, focusing on phase transitions that appear first order at the soft scale.

Note that, unlike the majority of studies utilising high-temperature dimensional reduction, we do not integrate out the temporal components of gauge fields to construct a \textit{softer} scale EFT~\cite{Kajantie:1995dw, Niemi:2021qvp}. The \textit{softer} scale EFT is only valid for sufficiently weak transitions, though its region of validity is expected to be relatively large~\cite{Kajantie:1995dw}. We have performed a brief comparison to this approach and find close agreement.

%%%%%%%%%%% Our EFT %%%%%%%%%%%
\subsubsection{The effective field theory} \label{sec:xsm_eft}
Given the power-counting assumptions of Section~\ref{sec:power_counting}, deriving the corresponding soft-scale EFT is standard, utilising from the imaginary time formalism of thermal field theory. The EFT is defined in 3d, being for the zero Matsubara modes, and the construction is called high-temperature dimensional reduction~\cite{Kajantie:1995dw}.

The 3d effective Lagrangian for the xSM reads
\begin{align} \label{eq:lagrangian_eft}
    \mathscr{L}_{3}             & = \mathscr{L}_\text{SM,3} + \mathscr{L}_{s,3} + \mathscr{L}_\text{portal,3},                                                  \\
    \mathscr{L}_{s,3}           & = \frac{1}{2}\partial_i s \partial_i s + b_{1,3}s + \frac{1}{2}m_{s,3}^2 s^2 + \frac{1}{3}b_{3,3}s^3 + \frac{1}{4}b_{4,3}s^4, \\
    \mathscr{L}_\text{portal,3} & = \frac{1}{2}a_{1,3}s \phi^\dagger \phi + \frac{1}{2}a_{2,3}s^2 \phi^\dagger \phi,
\end{align}
where we have used the same notation for the 3d effective scalar fields as the corresponding 4d fields. Note however that we canonically normalise the fields in 3d by scaling by appropriate powers of $T$ (plus perturbative corrections), so that
\begin{equation}
    S_3 = \int \mathrm{d}^3 x \mathscr{L}_{3}
\end{equation}
is dimensionless, and the 3d fields have mass dimension 1/2.
Expressions for the 3d effective parameters in terms of the 4d parameters of the original Lagrangian are known as matching relations. For this model, they have been derived in refs.~\cite{Brauner:2016fla, Schicho:2021gca, Niemi:2021qvp}.

The matching relations for this EFT can be derived as an expansion in integer powers of $x^2$, as only the hard scale is involved~\cite{Kajantie:1995dw}, and the relevant parameters are known up to one subleading power of $x^2$. Reaching this accuracy requires a two-loop computation and sets a yardstick for what follows; as far as possible, we aim to perform our analysis to the same accuracy. Higher dimension operators, and other operators not listed here, give contributions to the scalar effective potential that are further subleading; see refs.~\cite{Chala:2024xll, Niemi:2024vzw} for recent numerical investigations of higher-dimensional operators.

%%%%%%%%%%% Analysis within the EFT %%%%%%%%%%%
\subsection{Analysing phase transitions} \label{sec:analysing_phase_transitions}

The xSM admits a wide variety of possible cosmological thermal histories, which can be discerned by following the free energy and the scalar field expectation values, as functions of temperature. Given our motivation in terms of making predictions of gravitational wave signals, we focus on first-order phase transitions.

The pipeline from Lagrangian parameters to gravitational wave spectrum involves four largely separate steps~\cite{Caprini:2019egz, Caprini:2024hue}.%
\footnote{Interestingly, two recent works have carried out the whole prediction pipeline for a given microscopic Lagrangian~\cite{Wang:2024slx, Tian:2024ysd}.}
First, parameters of the phase transition are computed from the microscopic Lagrangian, using thermal quantum field theory. This first step is the main focus of our study. Second, the phase transition parameters are passed as input to macroscopic, classical calculations of the phase transition dynamics and resulting gravitational wave signal. Third, this signal is redshifted to today, based on a cosmological model. Finally, the redshifted signal is compared to the sensitivities of current and proposed experiments.

For the second step in the pipeline, we make use of fits to the gravitational waves spectra from large-scale lattice simulations of the macroscopic dynamics of bubble collisions and fluid flows~\cite{Hindmarsh:2017gnf, Jinno:2022mie}, as well as modelling of the evolution of a single bubble for given phase transition parameters~\cite{Giese:2020znk, Ai:2023see}.
In total, the fits for the gravitational wave density parameter $\Omega_\text{GW}$ per logarithmic frequency $f_{GW}$ take the form
\begin{align}
    \Omega_\text{GW}(f_{GW}) = F\left(f_{GW}; \{\text{phase transition parameters}\}\right),
\end{align}
where $F$ is a complicated fit function, and its arguments are the required phase transition parameters, outlined in Section~\ref{sec:phase_transition_parameters} below.

As mentioned, our main focus is on how gravitational wave predictions are affected by uncertainties in the phase transition parameters. So, we take the fit function $F$ as fixed, and simply propagate our uncertainties in the phase transition parameters through to the gravitational wave spectrum. See refs.~\cite{Guo:2021qcq,Athron:2023rfq} for complementary studies of the latter steps of the prediction pipeline.

%%%%%%%%%%% Phase transition parameters %%%%%%%%%%%
\subsubsection{Phase transition parameters} \label{sec:phase_transition_parameters}

For a first-order phase transition, perhaps the simplest observable is the critical temperature, $T_\text{c}$. This is the temperature at which two phases have the same free energy density. It depends only on homogeneous, equilibrium states, and consequently reaching higher perturbative orders is relatively straightforward.

For the soft-scale EFT of the xSM, the free energy density for given scalar background fields $\phi=\frac{1}{\sqrt{2}}(0, v)$ and $s=s_0$ is related to the 3d effective potential by
\begin{equation}
    f = T V_\text{3,eff}(v, s_0; T).
\end{equation}
Within perturbation theory, the free energy density of a phase $A$ can be evaluated by solving perturbatively for the minima of the potential, expanding around the leading order solution $v=v_A^{(0)}$, $s_0=s_{A}^{(0)}$. Here we denote the perturbative order by the superscript number in parentheses. Such an expansion yields order-by-order gauge invariant results for the free energy density in a given phase~\cite{Nielsen:1975fs, Fukuda:1975di, Laine:1994zq}, and consequently for the critical temperature,
\begin{equation}
    f_A(T_\text{c}) - f_B(T_\text{c}) = 0,
\end{equation}
here between phases $A$ and $B$.
% \begin{align}
%     f_A(T) &= T V^{(0)}_\text{3,eff}(v_A^{(0)}, s_{A}^{(0)}; T)
%     + T V^{(1)}_\text{3,eff}(v_A^{(0)}, s_{A}^{(0)}; T) + \dots
%     % \nonumber \\
%     % &\quad
%     % + T \left[
%     %     V^{(2)}_\text{3,eff}(v_A^{(0)}, s_{A}^{(0)}; T)
%     %     +\partial_{v_A} V^{(1)}_\text{3,eff}(v_A^{(0)}, s_{A}^{(0)}; T)v_A^{(1)}
%     %     +V^{(2)}_\text{3,eff}(v_A^{(0)}, s_{A}^{(0)}; T)
%     % \right].
% \end{align}

However, gravitational waves are not produced by homogeneous equilibrium states, so the critical temperature does not directly enter the gravitational wave spectrum. Instead the relevant temperature is that at which most gravitational waves are produced. This depends on the time evolution of the transition, and hence on the rate of bubble nucleation per unit volume, $\Gamma$. Assuming the Hubble expansion $H$ is radiation dominated, that the rate of bubble nucleation is exponentially sensitive to the temperature, and that the bubble size at collision is much larger than their initial size, one can derive the following approximate expression for the temperature $T_*$ at which the bubble wall area changes most rapidly,
\begin{align} \label{eq:nucleation_temperature}
    \left[-\log(\Gamma / H^4) + 4\log(\beta/H) - \log(8\pi v_w^3) + \log(-\log(f_\text{meta}))\right]_{T=T_*} = 0,
\end{align}
where $v_\text{w}$ is the asymptotic bubble wall speed, and $f_\text{meta}\approx 1/e$ is the fraction of the universe in the metastable state at this temperature~\cite{Enqvist:1991xw}, and we have defined
\begin{align}
    \beta = -H T \partial_T \log\Gamma.
\end{align}
It is possible to relax the assumption that the bubble nucleation rate is exponentially sensitive to the temperature, and the approximation of equation~\eqref{eq:nucleation_temperature} corresponds to \textit{moderate diligence} in the nomenclature of ref.~\cite{Guo:2021qcq}.

In addition to the phase transition parameters appearing in equation~\eqref{eq:nucleation_temperature}, there are two relevant phase transition parameters that characterise the change in energy density between the metastable and stable phases~\cite{Giese:2020znk, Ai:2023see},
\begin{align} \label{eq:alpha_psi}
    \alpha_* & = \frac{1}{3w_\text{meta}}\left(De + \frac{Df}{c_\text{s,stable}^2}\right), &
    \Psi_*   & = \frac{w_\text{stable}}{w_\text{meta}},
\end{align}
where all is evaluated at $T=T_*$, and $D X \equiv X_\text{meta} - X_\text{stable}$. The internal energy density $e$, the enthalpy density $w$, and the speed of sound $c_\text{s}$ can be derived from the free energy density via the usual thermodynamic relations. See ref.~\cite{Tenkanen:2022tly} for a study of the effects of higher order corrections to the speed of sound.
In effect, $\alpha_*$ determines the energy density released in the transition relative to that of the metastable phase, and $\Psi_*$ determines the ratio of the numbers of degrees of freedom in each the phase.

Given the four phase transition parameters: $T_*$, $\alpha_*$, $\Psi_*$, $\beta/H_*$, the bubble wall speed $v_w$ and the kinetic energy fraction $K$~\cite{Caprini:2019egz} can be determined in the local thermal equilibrium approximation. This neglects contributions to the friction on the bubble wall from out-of-equilibrium particles, but appears to be a reasonable approximation for the xSM, at least regarding the contribution of the top quark, which is expected to dominate the friction on the bubble wall~\cite{Laurent:2022jrs}. We make use of the fit formula in ref.~\cite{Ai:2023see}, valid for phase transitions during radiation domination.

Given these phase transition parameters, we use the gravitational wave spectrum fit of refs.~\cite{Hindmarsh:2015qta, Hindmarsh:2017gnf, Caprini:2019egz}, and account for Hubble expansion during the transition following ref.~\cite{Guo:2020grp}. In total, our analysis of the macroscopic dynamics of bubble collisions and gravitational wave production lies between the moderate and high diligence of ref.~\cite{Guo:2021qcq}. Our analysis of the microscopic thermal physics of the phase transition parameters is nevertheless state of the art, details of which follow in Section~\ref{sec:factorisation}.

%%%%%%%%%%% Factorisation %%%%%%%%%%%
\subsubsection{Factorising the microscopic computations} \label{sec:factorisation}

With the effective Lagrangian in hand~\eqref{eq:lagrangian_eft}, the next step is to analyse the phase transitions using it. The EFT is sensitive to the parameters and temperature of the full theory only through the parameters of its effective Lagrangian. This allows to factorise the computation of physical observables, and thereby identify the sources of different contributions.

To outline the analysis procedure, while keeping the formulae manageable, let us consider a phase transition in the singlet direction, and ignore for the moment the Higgs and other fields within the EFT.
The tree-level potential for the singlet background field $s_0$ is
\begin{equation} \label{eq:singlet_Vtree}
    V_3^{(0)}(s_0) = b_{1,3}s_0 + \frac{1}{2}m_{s,3}^2 s_0^2 + \frac{1}{3}b_{3,3}s_0^3 + \frac{1}{4}b_{4,3}s_0^4.
\end{equation}
Here the superscript label ${}^{(0)}$ denotes the loop order within the EFT. This is independent of the order at which the effective parameters were matched.

To analyse phase transitions in this potential, it is convenient to shift the background field by a constant to remove the cubic term, $s_0 \to s_0 - b_3/(3 b_4)$,
\begin{equation} \label{eq:singlet_Vtree_shifted}
    V_3^{(0)}(s_0) = \tilde{b}_{1,3}s_0 + \frac{1}{2}\tilde{m}_{s,3}^2s_0^2  + \frac{1}{4}b_{4,3}s_0^4,
\end{equation}
where we have defined the shifted parameters
\begin{align}
    \tilde{b}_{1,3}   & = b_{1,3} + \frac{2 b_{3,3}^3}{27 b_{4,3}^2} - \frac{m_{s,3}^2 b_{3,3}}{3 b_{4,3}}, &
    \tilde{m}_{s,3}^2 & = m_{s,3}^2 - \frac{b_{3,3}^2}{3 b_{4,3}}.
\end{align}
Given this shifted potential, a line of phase transitions can be identified at $\tilde{b}_{1,3}=0$, based on an enhanced $Z_2$ symmetry. At this point, and assuming $\tilde{m}_{s,3}^2<0$ so that the transition appears first order, there are two minima for the singlet background field at
\begin{equation}
    s_0^{(0)} = \pm \sqrt{\frac{-\tilde{m}_{s,3}^2}{b_{4,3}}}.
\end{equation}
The jump in the background field value at the transition is then $\Delta \left<s\right>^{(0)} = 2 (-\tilde{m}_{s,3}^2/b_{4,3})^{1/2}$.

Within the EFT, the one- and two-loop corrections to the potential take the form,
\begin{align} \label{eq:singlet_Vone-loop}
    V_3^{(1)}(s_0) & =
    -\frac{1}{3(4\pi)}M_3^2(s_0)^{3/2}, \\
    V_3^{(2)}(s_0) & =
    \frac{1}{(4\pi)^2} \left[\frac{3}{4}b_{4,3} M_3^2 + 12b_{4,3}^2 s_0^2\left(\log\left(\frac{3 M_3}{\mu_3 }\right)-\frac{1}{2}\right) \right],
\end{align}
in terms of the field-dependent effective mass $M_3^2 \equiv \partial_{s_0}^2V_3^{(0)}$ and the renormalisation scale of the 3d EFT $\mu_3$.

The minima of the effective potential, and hence the jump in the background field at the transition, can then be expanded perturbatively as~\cite{Gould:2021dzl}
\begin{equation} \label{eq:delta_s}
    \frac{\Delta \left< s \right>}{\Delta \left< s \right>^{(0)}} =
    1
    + \frac{3b_{4,3}}{\sqrt{2}(4\pi)|\tilde{m}_{s,3}|}
    - \frac{3b_{4,3}^2}{4(4\pi)^2|\tilde{m}_{s,3}|}\left[1+16\log\left(\frac{3\sqrt{2}|\tilde{m}_{s,3}|}{\mu_3}\right)\right]
    +O(x^3),
\end{equation}
This is a strict loop expansion within the EFT, which ensures order-by-order gauge invariance~\cite{Laine:1994zq, Croon:2020cgk, Hirvonen:2021zej}. The expansion parameter is $b_{4,3}/(4\pi |\tilde{m}_{s,3}|)\sim x$, so that this gives an expansion in integer powers of $x$. Our power counting assumptions then generalise the structure of the expansion to the full xSM. This approach has also been adopted in refs.~\cite{Gould:2021oba, Ramsey-Musolf:2024ykk}.

Equation~\eqref{eq:delta_s} is written solely in terms of the 3d effective parameters. These parameters, in turn, can be written in terms of the original 4d parameters, as an expansion in integer powers of $x^2$. As a consequence, the full perturbative expansion for $\Delta \left< s \right>/\Delta \left< s \right>^{(0)}$ takes the schematic form
\begin{align} \label{eq:delta_s_expansion}
    \frac{\Delta \left< s \right>}{\Delta \left< s \right>^{(0)}} =
    \underbrace{\left(1+x^2+O(x^4)\right)}_\text{hard}
    \times
    \underbrace{\left(1+x+x^2+O(x^3)\right)}_\text{soft}
    \times
    \underbrace{\left(1+O(x^4)\right)}_\text{ultrasoft}.
\end{align}
Here we have also shown the magnitude of the leading corrections from the ultrasoft energy scale, $x^2 \pi T$. The contribution of the ultrasoft scale is subleading but nonperturbative, as all loop topologies at this scale contribute to the $O(x^4)$ term~\cite{Linde:1980ts, Kajantie:2002wa}.

The same form~\eqref{eq:delta_s_expansion} of expansion holds for any phase transition observable $\mathcal{O}_\text{hom}$ that only depends on homogeneous equilibrium field configurations, such as the critical temperature or the latent heat,
\begin{equation}
    \frac{\mathcal{O}_\text{hom}}{\mathcal{O}_\text{hom}^{(0)}} =
    \underbrace{\left(1+x^2+O(x^4)\right)}_\text{hard}
    \times
    \underbrace{\left(1+x+x^2+O(x^3)\right)}_\text{soft}
    \times
    \underbrace{\left(1+O(x^4)\right)}_\text{ultrasoft}.
\end{equation}
This equation summarises the accuracy of our equilibrium calculations, where the big $O$ notation shows the size of the leading errors arising from terms that we have not computed.

Note that large logarithms (i.e.\ logarithms of $x$) may arise in intermediate results, due to ratios of hard, soft and ultrasoft scales, but they can be removed by treating each scale separately. For each energy scale, we can use the renormalisation group equations of the corresponding EFT to introduce a separate renormalisation scale, ensuring sufficient freedom to give all logarithms $O(1)$ arguments~\cite{Weinberg:1980wa}. This is a key advantage of EFT.

The EFT approach to bubble nucleation~\cite{Gould:2021ccf} outlines how the same techniques can be applied to computing the rate of bubble nucleation. The first step is to identify the nucleating degrees of freedom, and the corresponding EFT. For the xSM these are the two scalar fields, which have effective masses at the soft scale. Therefore, the nucleation scale effective theory is simply the 3d EFT of Section~\ref{sec:xsm_eft} that we have been using thus far. The nucleation rate is then calculated via a saddlepoint approximation to a path integral for the EFT.

The rate of bubble nucleation is determined by the energy of a critical bubble configuration: a static, spherical saddlepoint of the nucleation scale effective action~\cite{Langer:1969bc, Affleck:1980ac, Linde:1981zj, Gould:2021ccf},
\begin{equation} \label{eq:saddlepoint}
    \frac{\mathrm{d}^2\varphi_a(r)}{\mathrm{d}r^2}
    + \frac{2}{r}\frac{\mathrm{d}\varphi_a(r)}{\mathrm{d}r}
    - \partial_a V_3^{(0)}(\varphi(r)) = 0.
\end{equation}
Here $r$ is a radial coordinate, and $a$ is an index which runs over the two scalar field backgrounds, $\varphi_a(r) = (v(r), s_0(r))$.

In the construction of our effective theory, a derivative expansion was used, based on the assumption that $\partial^2 /(\pi T)^2 \ll 1$. Due to equation~\eqref{eq:saddlepoint}, the critical bubble varies over distance scales of order the effective Compton wavelengths of the nucleating scalar fields. By assumption, the scalar fields have masses at the soft scale, $x \pi T$, so the derivative expansion for constructing the EFT is indeed justified, being an expansion in powers of $\partial^2\varphi /(\pi T)^2 \sim x^2\ll 1$.

Note that it is the tree-level potential of the soft-scale EFT that appears in equation~\eqref{eq:saddlepoint}, as follows from a saddlepoint approximation to the path integral~\cite{Gould:2021ccf}.
It would be inappropriate to replace this with a potential including loop corrections from within the EFT, such as $V_3^{(0)}\to V_3^{(0)}+V_3^{(1)}$; doing so would imply a breakdown of the derivative expansion, a double counting of degrees of freedom in the path integral, and would lead to unphysical imaginary parts, as discussed in refs.~\cite{Croon:2020cgk, Gould:2021ccf}. Instead, loop corrections from within the EFT must be carried out in the background of the inhomogeneous bubble configurations, including all orders in derivatives.

The logarithm of the nucleation rate takes the form
\begin{align} \label{eq:log_nucleation_rate}
    -\log\left(\Gamma/{T^4}\right) & = S_{3}[\varphi]
    -\log\left(A_\text{stat}/{T^3}\right) -\log\left(A_\text{dyn}/{T}\right), \\
                                   & =S_{3}[\varphi]\left(1 + O(x)\right).
\end{align}
Here $A_\text{stat}$ is the statistical prefactor and $A_\text{dyn}$ is the dynamical prefactor~\cite{Langer:1974cpa}. The statistical prefactor accounts for static fluctuations about the critical bubble and involves inhomogeneous but equilibrium physics. It can be written in terms of functional determinants of radial differential operators. The dynamical prefactor accounts for time-dependent fluctuations and involves both inhomogeneous and out-of-equilibrium physics. Over the past decades there has been no consensus on its precise form~\cite{Langer:1969bc, Affleck:1980ac, Linde:1981zj, Arnold:1987mh, Hanggi:1990zz}. Recent theoretical works have made progress in this regard, when the real-time dynamics admits a Langevin~\cite{Bodeker:1998hm, Ekstedt:2022tqk} or Boltzmann~\cite{Hirvonen:2024rfg} effective description, but discrepancies between analytical and numerical approaches remain~\cite{Pirvu:2024ova, Pirvu:2024nbe, Gould:2024chm}.

Given the computational and theoretical challenges of computing $A_\text{stat}$ and $A_\text{dyn}$, we do not do so in this paper. As such, the leading errors for our predicted nucleation rates are $O(x)$, much larger than those for the critical temperature $O(x^3)$. The following equation summarises the perturbative orders we reach for observables $\mathcal{O}_\text{nucl}$ directly involving the nucleation rate, such as $T_*$ and $\beta/H_*$,
\begin{equation}
    \frac{\mathcal{O}_\text{nucl}}{\mathcal{O}^{(0)}_\text{nucl}} =
    \underbrace{\left(1+x^2+O(x^4)\right)}_\text{hard}
    \times
    \underbrace{\left(1+O(x)\right)}_\text{soft}
    \times
    \underbrace{\left(1+O(x^4)\right)}_\text{ultrasoft}.
\end{equation}
Due to the factorisation of contributions from hard and soft scales, it is still worthwhile to include the $x^2$ corrections from the hard scale. To reach the same order for the soft-scale physics would require computing two-loop vacuum diagrams in the background of the critical bubble, and subleading corrections to the dynamical prefactor, both of which are beyond the current state of the art~\cite{Bezuglov:2018qpq, Ekstedt:2022tqk}.

The $O(x)$ uncertainty in $T_*$ infects every other phase transition parameter that enters the gravitational wave spectrum, all of which are evaluated at this temperature. For homogeneous equilibrium quantities such as $\alpha_*$, it is nevertheless possible to include some higher order corrections from the soft scale. Inspired by successful comparisons to lattice Monte-Carlo simulations in other models~\cite{Gould:2022ran, Gould:2023ovu}, we write the free energy density evaluated at $T_*$ as
\begin{equation}
    f(T_*) = f(T_\text{c}, \delta T_*),
\end{equation}
where $\delta T_* \equiv T_\text{c} - T_*$, and then expand both the functional form of $f$ and the argument $T_\text{c}$ to next-to-next-to-leading order, while $\delta T_*$ remains at leading order because higher orders are not known. In this way, we can include higher-order homogeneous equilibrium contributions while avoiding a breakdown of the expression from $\delta T_*$ changing sign. Note also that we apply this expansion to the physical quantities $f$, $e$ and $w$ separately before taking the combinations in equation~\eqref{eq:alpha_psi}, another example of factorisation, which we have found to give significant improvements over a strict expansion of $\alpha_*$.

\begin{table}
    \centering
    \begin{tabular}{|c|c|c|c|}
        \hline
                             & hard           & soft equilibrium & soft nonequilibrium                  \\
        \hline
        $T_\text{c}$         & $1+x^2+O(x^4)$ & $1+x+x^2+O(x^3)$ & ---                                  \\
        \hline
        $T_{*}, \beta/H_{*}$ & $1+x^2+O(x^4)$ & $1+O(x)$         & $1 + O(x)$                           \\
        \hline
        $\alpha_{*}$         & $1+x^2+O(x^4)$ & $1+x+x^2+O(x^3)$ & $1 + O(x)$                           \\
        \hline
        $\Psi_{*}$           & $1+x^2+O(x^4)$ & $1+x^2+O(x^3)$   & $1 + O(x)$                           \\
        \hline
        $v_\text{w}, K$      & $1+x^2+O(x^4)$ & $1+x+x^2+O(x^3)$ & $1 + O(x) + O(\delta f/f_\text{eq})$ \\
        \hline
    \end{tabular}
    \caption{Summary of perturbative orders reached for the phase transition parameters, broken down by energy scale. In each case 1 denotes the leading order, powers of $x$ denote subleading corrections that have been included, and the big $O$ symbol denotes the leading theoretical uncertainty. For the soft scale we also separate those contributions that can be accounted for in a homogeneous equilibrium phase, and those which cannot. In all cases, we rely on the assumptions of Section~\ref{sec:power_counting}.\label{table:orders}}
\end{table}

For the bubble wall speed, taking the results of the above computations as input, we adopt the local equilibrium approximation~\cite{Konstandin:2010dm, BarrosoMancha:2020fay, Balaji:2020yrx, Ai:2021kak, Ai:2023see}. This is the leading order of the Chapman-Enskog expansion~\cite{chapman1952mathematical, Dashko:2024spj}. It amounts to neglecting deviations from equilibrium for distribution functions in the Boltzmann equations about the growing bubble wall, and hence we expect the expansion parameter to be $O(\delta f/f_\text{eq})$ .

A more complete state-of-the-art calculation of the bubble wall speed~\cite{Laurent:2022jrs, DeCurtis:2024hvh, Dashko:2024spj}, including the solution of the Boltzmann equations for the out-of-equilibrium distribution functions, would reduce these errors in the bubble wall speed to $O(1/|\log(x)|)$, where the residual uncertainty is due to the use of the leading-logarithm approximation for the collision integrals.

Table~\ref{table:orders} summarises the orders reached in the present paper. From this it is clear that the leading residual errors stem from nonequilibrium physics related to bubble nucleation and growth.

An outline of our numerical implementation is given in Appendix~\ref{appendix:numerics}.

%%%%%%%%%%%%% Results %%%%%%%%%%%%%
\section{Results} \label{sec:results}

In what follows, we separate predictions based on the powers of $x$ that they include from the weak-coupling expansion, into leading order (LO), next-to-leading order (NLO) and next-to-next-to-leading order (NNLO),
\begin{align}
    \frac{\mathcal{O}}{\mathcal{O}^{(0)}} = \underbrace{\overbrace{\underbrace{1}_\text{LO} +\ x}^\text{NLO}\ +\ x^2}_\text{NNLO}\ +\ O(x^3).
\end{align}
Note, however, that we do not break the factorisation by energy scales of our results.
For computations involving inhomogeneous and nonequilibrium physics, our computations are incomplete at higher orders, so we denote the predictions as partial next-to-leading order (NLO$^*$) and partial next-to-next-to-leading order (NNLO$^*$).

To estimate the intrinsic uncertainty in predictions, we use the residual dependence on the renormalisation scale as a proxy. To do so we solve \textit{exactly} the one-loop (two-loop) renormalisation group equations for the 4d (3d) Lagrangian parameters. This gives all running parameters with relative accuracy to orders $1+x^2$ and with $O(x^4)$ errors. The renormalisation scale dependence of physical predictions should cancel up to the order computed, with uncancelled higher orders remaining and depending on the choice of renormalisation scales.

Concretely, the pair of 3d and 4d renormalisation scales $(\mu_3/T, \mu/(\pi T))$ are chosen to take values in the set $\{(1, 1), (0.5, 1), (2, 1), (1, 0.5), (1, 2)\}$.%
\footnote{These choices are conventional~\cite{Kainulainen:2019kyp, Gould:2021oba}. However, with hindsight somewhat smaller values for $\mu_3/T$ would have been preferable, as the convergence of perturbation theory was better at $\mu_3/T=0.5$ than at $1$ or $2$.}
Following ref.~\cite{Croon:2020cgk}, the relative uncertainty in a prediction $X$ is then defined as
\begin{align}
    \frac{\Delta X}{X} \equiv \frac{\max(X) - \min(X)}{\min(X)},
\end{align}
where the maximum and minimum are taken over the set of pairs of renormalisation scales. As the default prediction, we take the element $(\mu_3/T, \mu/(\pi T))=(1,1)$.

\subsection{Parameter scan} \label{sec:parameter_scan}

The xSM has 5 independent Lagrangian parameters in addition to those of the Standard Model, as one of the 6 apparent parameters can be removed by a choice of field origin for $s$~\cite{Espinosa:2011ax}. There is a special subset of parameter choices that admits the $Z_2$ symmetry $s\to -s$ and hence is radiatively stable. It satisfies $a_1=b_1=b_3=0$ and hence has only 3 independent parameters.

We choose to study the generic non-$Z_2$ symmetric parameter space. We avoid the $Z_2$ symmetric subset, as this may lead to domain wall formation, which requires separate consideration~\cite{Blasi:2022woz, Blasi:2023rqi, Agrawal:2023cgp}.

We perform a random scan of the parameter space, as this provides better coverage of multidimensional parameter spaces than a uniform scan~\cite{AbdusSalam:2020rdj}. Our particular choice of parameter ranges largely follow ref.~\cite{Niemi:2024vzw}, though we limit to somewhat smaller four-point couplings and singlet masses. Larger singlet masses typically require larger portal couplings to significantly modify the electroweak phase transition~\cite{Ramsey-Musolf:2019lsf}. Specifically, input parameter points are drawn from the following distributions
\begin{align}
    M_s        & \in \mathcal{U}(10, 700)~\text{GeV},   \\
    \sin\theta & \in \mathcal{U}(-0.25, 0.25),          \\\
    a_2        & \in \mathcal{U}(-0.5, 4),              \\
    b_3        & \in \mathcal{U}(-300, 300)~\text{GeV}, \\
    b_4        & \in \mathcal{U}(0.001, 2),
\end{align}
where $\mathcal{U}(a,b)$ is the uniform distribution on $[a, b)$. The parameter $M_s$ is the physical singlet (pole) mass, and is matched to the corresponding \MSbar\ parameter $m_s$ at one-loop order, along with the Standard Model experimental input parameters; for details see ref.~\cite{Niemi:2021qvp}. The other singlet parameters are input directly as \MSbar\ parameters at the input renormalisation scale, which we take as the $Z$-pole mass, $\mu_\text{input}=M_Z$.

Note that these parameter ranges approximate current exclusion limits from collider searches, in particular the constraint on the mixing angle $|\sin\theta|\leq 0.25$. See refs.~\cite{Alves:2018jsw, Ramsey-Musolf:2024ykk, Niemi:2024axp, Aboudonia:2024frg} for further details and studies of the complementarity between collider and gravitational wave experiments.

The three and four-point couplings are then further constrained by perturbative unitarity, from $\mu_\text{input}$ up to renormalisation scales of at least $\mu=10$~TeV. We also ensure the stability of the electroweak vacuum at zero temperature and at one-loop order (at least for minima around or below the electroweak scale).

In total, we analyse $10^5$ benchmark parameter points satisfying these conditions. The complete list can be found as comma separated values at ref.~\cite{gould_2024_14031507}.

\subsection{Thermal histories} \label{sec:histories}

Of our $10^5$ parameter points, we find approximately 9/10 have Standard-Model-like single-step symmetry-breaking phase transitions, and 1/10 have alternative thermal histories involving first-order phase transitions. This conclusion depends sensitively both on our choice of parameter distributions, and on our assumption that the transition takes place at the soft scale. It nevertheless supports the expectation that a minimal modification of the Standard Model will often lead to a Standard-Model-like thermal history.

Note that within the soft-scale EFT, these single-step symmetry-breaking phase transitions appear to be of second order, and hence the Higgs scalar in fact lives at lower energies still than the soft scale. A corresponding lower energy EFT is required to study these transitions. Further, identifying the borderline between first-order, second-order and crossover transitions requires nonperturbative input~\cite{Gould:2019qek, Ramsey-Musolf:2024ykk}.

Of the thermal histories involving first-order phase transitions at the soft scale, approximately 1/4 do not complete (i.e.\ the bubbles do not percolate), so that the end state is not the electroweak symmetry-breaking phase in contradiction with observations. Thus, we are left with around 7000 parameter points for further study.

Note that for a small fraction of the remaining parameter points, large corrections at higher orders yield unphysical predictions such as negative $\alpha_*$. In what follows, at each perturbative order and for each observable we have chosen to focus on those parameter points that yield physically sensible predictions. It may be that alternative EFTs or even lattice Monte-Carlo simulations are needed to resolve the seemingly nonperturbative points, but we leave this question for future work. Similar behaviours have been met and studied in ref.~\cite{Papaefstathiou:2020iag}, which studied how the existence of a phase transition depends on the renormalisation scale, and ref.~\cite{Gould:2023jbz} which found $O(1)$ changes in phase transition parameters between perturbative orders for many parameter points in a Yukawa model.

\subsection{Critical temperature} \label{sec:Tc}

\begin{figure}[t]
    \centering
    \includegraphics[width=0.48\textwidth]{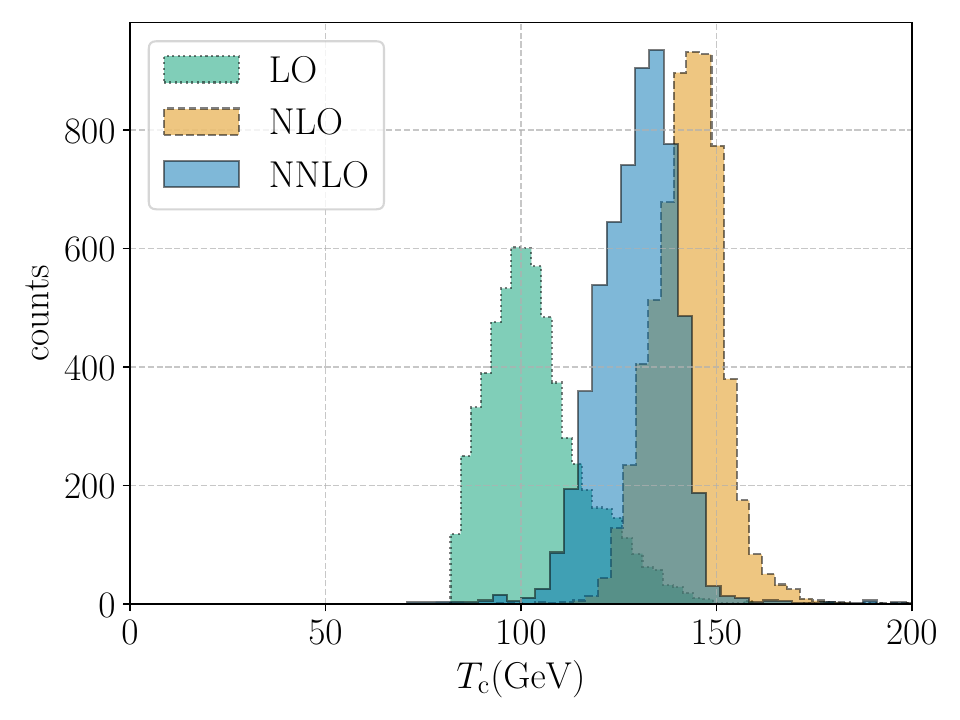}
    \includegraphics[width=0.48\textwidth]{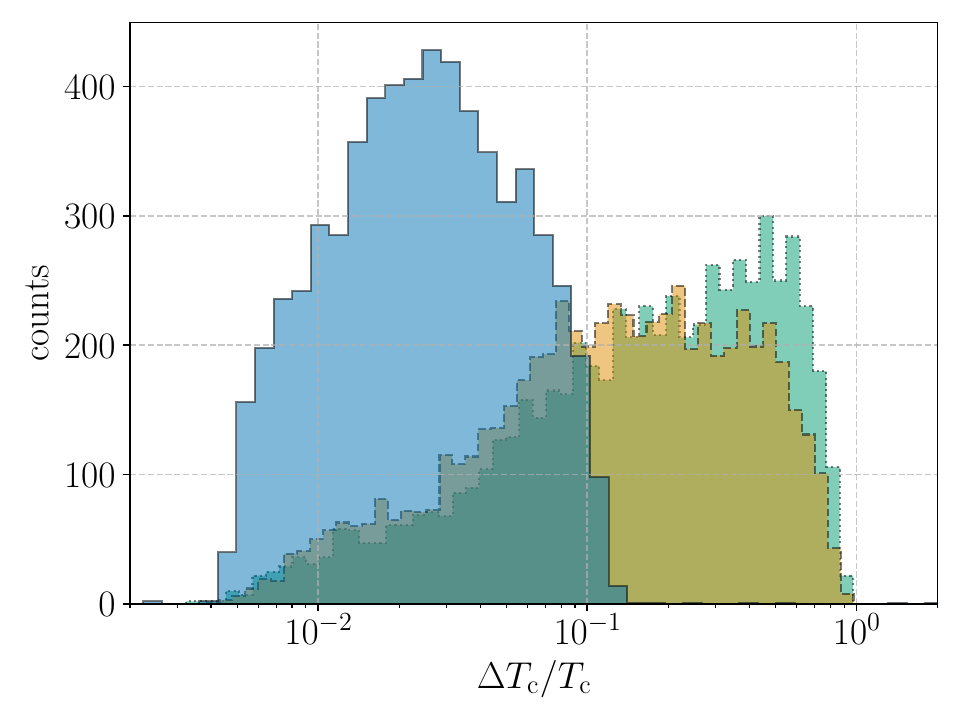}
    \caption{Histograms of the critical temperature (left) and its relative error (right) for a scan over the xSM. Shown are the results at LO, NLO and NNLO in the coupling expansion, all for the same physical parameter points.}
    \label{fig:histogram_Tc}
\end{figure}

Figure~\ref{fig:histogram_Tc} shows our results for the critical temperature. The left hand plot shows a histogram of the critical temperatures for our scan computed at each perturbative order. All three distributions contain approximately 7000 parameter points. Perhaps surprisingly, the distributions at each order are relatively distinct, so that the perturbative order of the calculation determines the critical temperature more strongly than the physical parameter point. Nevertheless, by eye one can see that the NLO and NNLO distributions overlap more than the LO and NLO distributions, showing apparent convergence of the expansion.

The right hand plot shows a histogram of the relative theoretical uncertainties in the prediction at each perturbative order.
At LO and NLO the relative uncertainty in $T_\text{c}$ is typically between 0.1 and 1, while at NNLO it is reduced to typically between 0.01 and 0.1. This is a significant reduction in uncertainty at NNLO, and a strong indication that our perturbative expansion is converging for this observable.

The theoretical uncertainty, as measured by residual renormalisation scale dependence, is not reduced at NLO with respect to LO. The first reduction in renormalisation scale dependence is at NNLO. This result is expected because NLO is only smaller than LO by one power of $x$, yet renormalisation group running appears at $O(x^2)$ and hence at NNLO~\cite{Gould:2021oba}. Figure~\ref{fig:histogram_Tc} demonstrates the importance of NNLO corrections for making accurate predictions at the whole-model level.

\subsection{Gravitational waves} \label{sec:gravitational_waves}

The gravitational wave spectrum is a much more complicated observable than the critical temperature, involving inhomogeneous and out-of-equilibrium physics. It is also very sensitive to the values of phase transition parameters, in particular depending on the percolation temperature $T_*$ as strongly as a high inverse power~\cite{Croon:2020cgk}. The relative uncertainties are therefore expected to be significantly larger than for the critical temperature.

\begin{figure}[t]
    \centering
    \includegraphics[width=0.48\textwidth]{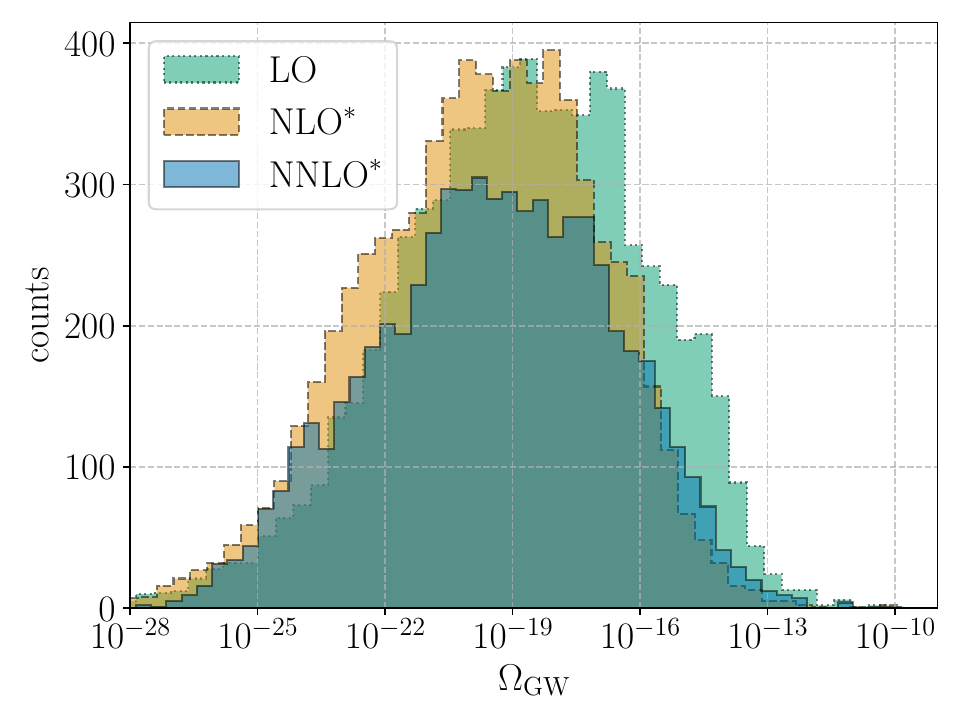}
    \includegraphics[width=0.48\textwidth]{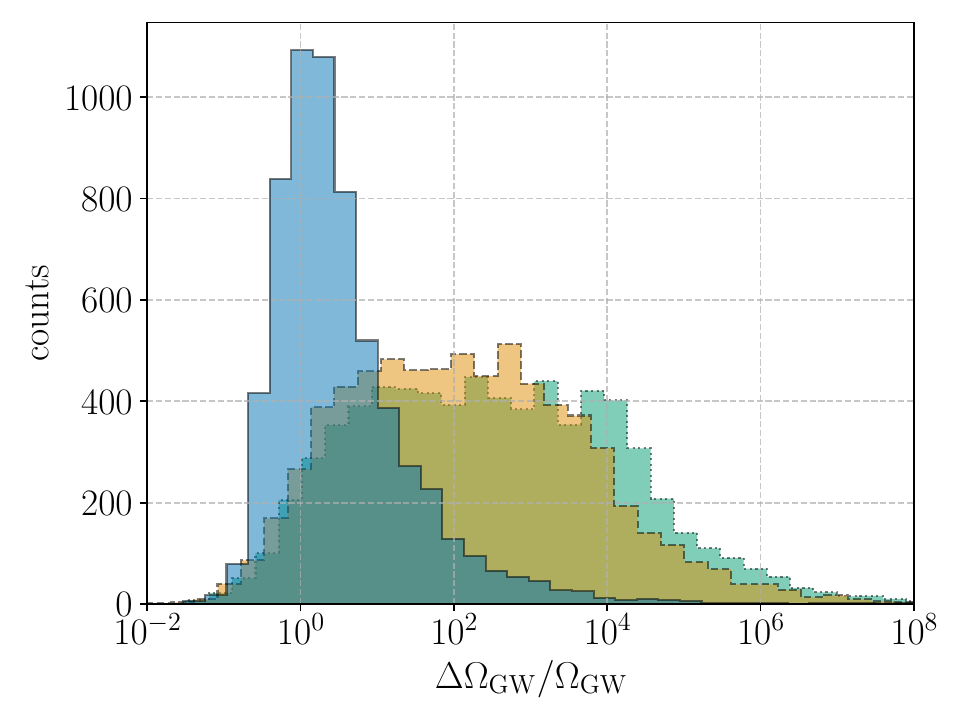}
    \caption{Histograms of the gravitational wave peak amplitude (left) and its relative error (right) for a scan over the xSM. Shown are the results at LO, partial next-to-leading order (NLO$^*$) and partial next-to-next-leading order (NNLO$^*$) in the coupling expansion, all for the same physical parameter points..}
    \label{fig:histogram_gws}
\end{figure}

Figure~\ref{fig:histogram_gws} shows histograms of the gravitational wave peak amplitude and of its uncertainty. We remind the reader that our higher-order results for this observable are partial, in that they do not include higher-order corrections arising from inhomogeneous and out-of-equilibrium physics, indicated by the superscript $^*$.

The left hand plot of figure~\ref{fig:histogram_gws} shows a very wide spectrum of predictions for the gravitational wave amplitude, with the bulk lying in the range $10^{-27}$ to $10^{-13}$. All three orders broadly agree on this range. Nevertheless, the strongest transitions at LO are missing at higher orders. In other words, the strongest transitions appear to be nonperturbative. Further work is needed to obtain reliable predictions for these parameter points. Nevertheless, the distribution of points at NNLO$*$ is shifted to somewhat higher values than at NLO$^*$, a feature consistent with many previous studies which have found that two-loop order corrections often yield stronger phase transitions~\cite{Kainulainen:2019kyp, Niemi:2021qvp, Niemi:2024vzw}.

The right hand plot of figure~\ref{fig:histogram_gws} shows how the relative uncertainty in the gravitational wave amplitude depends on the order of the calculation, and the results are rather striking. At LO and NLO there is a very wide spread of relative uncertainties up to very large values, being almost flat between $10$ and $10^4$. As expected, the NNLO$^*$ predictions have much smaller uncertainties on average than those at LO and NLO$^*$, with the bulk of the points lying between about 0.5 and 50. The distribution of relative uncertainties at NNLO$^*$ is also much more strongly peaked, with the peak around 1--2.

\begin{figure}[t]
    \centering
    \includegraphics[width=\textwidth]{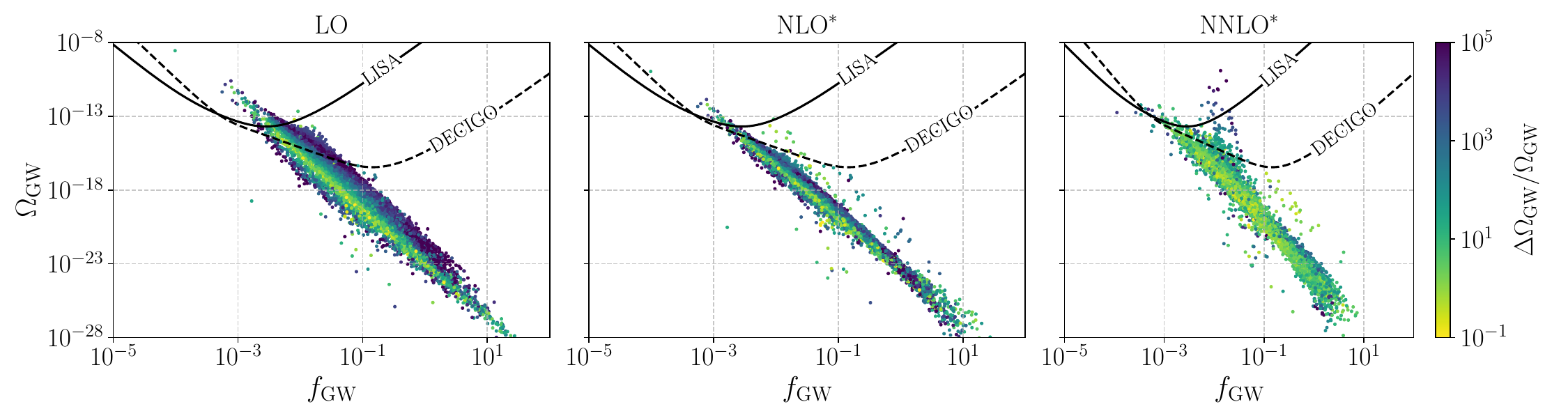}
    \caption{Results for the gravitational wave peak amplitude and frequency, together with peak-integrated sensitivity curves from ref.~\cite{Schmitz:2020syl}. The colour reflects the magnitude of the theoretical uncertainty in the peak amplitude, and hence the yellower points at NNLO$^*$ are more accurate. The calculations at each order are for the same set of physical parameter points.}
    \label{fig:scatter_gws}
\end{figure}

In figure~\ref{fig:scatter_gws} the peak gravitational wave signals are plotted as points on the frequency-amplitude plane, together with peak-integrated sensitivity curves from ref.~\cite{Schmitz:2020syl} for the proposed LISA~\cite{LISA:2017pwj} and DECIGO~\cite{Kawamura:2011zz} experiments. The colour of each point indicates the magnitude of the theoretical uncertainty in the peak amplitude.

In all three calculations, the peak amplitudes form a band from top left to bottom right, a quasi-universal feature of gravitational wave signals of electroweak phase transitions~\cite{Eichhorn:2020upj}. At LO, and to a lesser extent NLO$^*$, there is a clear geometric structure to the distribution of uncertainties on the plane, with larger uncertainties at points further from a central midrib. At NNLO$^*$ this geometric structure is washed out, and the majority of points display smaller uncertainties. Nevertheless, there are a number of outliers, in particular at large amplitudes, with large uncertainties even at NNLO$^*$.

Figure~\ref{fig:scatter_gws} clearly reveals the degrees of convergence of the perturbative expansion. For the bulk of the band, perturbation theory appears to be converging relatively well, with the spread of points remaining somewhat static and the uncertainty reducing. However, for those points leading to the strongest signals, in particular those well inside the LISA sensitivity band at LO, perturbation theory appears not to be converging.

%%%%%%%%%%%%% Conclusions %%%%%%%%%%%%%
\section{Conclusions} \label{sec:conclusions}

In the present paper we have carried out a state-of-the-art perturbative study of the phase transitions and corresponding gravitational wave signals of the real-scalar extended Standard Model. We have worked to NNLO, or two-loop order, in all cases where an expansion around a homogeneous, equilibrium background was possible, and to lower orders for inhomogeneous and out-of-equilibrium physics. We have performed a large-scale scan of the 5-dimensional parameter space of the model, and investigated the convergence of perturbative predictions for the critical temperature and gravitational wave signal, complementing several recent studies~\cite{Niemi:2024vzw, Niemi:2024axp, Lewicki:2024xan, Ramsey-Musolf:2024ykk}.

For the critical temperature, the weak-coupling expansion appears to converge well. Curiously, the distribution of results are distinct at each perturbative order. In agreement with the arguments of ref.~\cite{Gould:2021oba}, we find that theoretical uncertainties are comparable between LO and NLO, and are first reduced only at NNLO, by around one order of magnitude. This underscores the importance of reaching NNLO for studies of cosmological phase transitions.

For the gravitational wave amplitude, the broad brushstrokes of the distribution of predictions agree between orders. Yet relative uncertainties are huge at LO and NLO$^*$. Most points in our parameter scan have relative uncertainties at LO and NLO$^*$ between 1--$10^5$, with the bulk lying between 10--$10^4$. At NNLO$^*$ predictions are again significantly improved, with typical relative uncertainties being $O(1)$.
Parameter points with relatively weak gravitational wave signals are described well by the weak-coupling expansion, inspiring confidence in these predictions.
However, the status of those parameter points with the strongest gravitational wave signals is less clear. The convergence of the weak-coupling expansion is particularly poor for those points within the sensitivity ranges of future gravitational wave detectors LISA and DECIGO.

How then to make reliable predictions for the strongest transitions? The example of the critical temperature, which appears perturbatively under control for the entire parameter scan, suggests that a complete NNLO calculation should suffice. Hence, the incompleteness of our NNLO$^*$ computation with regard to the inhomogeneous and out-of-equilibrium physics may be at fault. This motivates higher-order computations of the bubble nucleation rate and bubble wall speed, both serious theoretical and computational challenges. For the bubble nucleation rate, a complete NNLO computation would require evaluating two-loop vacuum integrals using Green functions solved in the background of the critical bubble~\cite{Ekstedt:2022tqk, Bezuglov:2018qpq}, as well as resolving open theoretical questions related to the dynamical prefactor~\cite{Hirvonen:2024rfg, Pirvu:2024ova, Pirvu:2024nbe, Gould:2024chm}. Perhaps even harder would be a complete NNLO computation of the bubble wall speed, as the current state of the art does not even compute the full LO, relying on a leading logarithm approximation for collision integrals in the Boltzmann equations~\cite{Laurent:2022jrs}.

Alternatively, it may be that our power-counting assumptions are not applicable for the strongest transitions. In particular, the key assumption that the transitioning fields have effective masses at the soft scale, $x\pi T$, may fail for very strong transitions where the jump in scalar background expectation values is parametrically larger than $T$. In this case higher dimension operators become important~\cite{Chala:2024xll}, and the construction of the EFT requires amending. Alternative power-counting assumptions for such a possibility have been proposed in refs.~\cite{Lofgren:2023sep, Gould:2023ovu, Kierkla:2023von, Camargo-Molina:2024sde}.

Our main focus has been the convergence of perturbative quantum field theory for the parameters describing a cosmological first-order phase transition. Yet, even with perfect predictions for these parameters, there will still be theoretical uncertainties from our imperfect understanding of the classical and hydrodynamic evolution of the transition~\cite{Auclair:2022jod, Dahl:2024eup, Caprini:2024gyk}, as well as experimental uncertainties from foreground astrophysical sources of gravitational waves~\cite{Hindmarsh:2024ttn}. To maximise what we can learn about the thermal history of the early universe from future observations requires a concerted scientific effort across the complete pipeline from Lagrangian parameters to experimental data. As demonstrated here, LO and NLO$^*$ prediction uncertainties for given Lagrangian parameters are huge, and potentially the largest uncertainties in whole the pipeline~\cite{Lewicki:2024xan}. This underscores the importance of further developing the microscopic theory of cosmological phase transitions.

%%%%%%%%%%%%% Acknowledgemends and end matter %%%%%%%%%%%%%
\acknowledgments
O.G.\ would like to acknowledge helpful conversations with
A.~Ekstedt,
M.~Lewicki,
L.~Niemi,
P.~Schicho
and T.V.I.~Tenkanen.
O.G.\ would also like to thank L.~Niemi and T.V.I.~Tenkanen for sharing their Mathematica implementation for the xSM, and T.V.I.~Tenkanen for his insightful comments on the manuscript.
O.G.\ was supported by a Royal Society Dorothy Hodgkin Fellowship.
P.M.S.\ was supported by STFC consolidated grant number ST/T000732/1.
We are grateful for access to the University of Nottingham's Ada HPC service.

\section*{Data availability statement}
The code used for this project is open source and available at ref.~\cite{cosmoxsm}. The parameter scan and results are collected as plain text files of comma-separated values, at ref.~\cite{gould_2024_14031507}.

%%%%%%%%%%%%% Appendix %%%%%%%%%%%%%
\appendix

%%%%%%%%%%%%% Formulae %%%%%%%%%%%%%
\section{Implementation details} \label{appendix:implementation}

Here we give some practical details of our calculation.

%%%%%%%%%%%%% Feynman diagrams %%%%%%%%%%%%%
\subsection{Feynman diagrammatics} \label{appendix:feynman_diagrams}

At NNLO we require the two-loop effective potential within the soft-scale EFT. We include all effects of the temporal components of gauge fields directly in the potential, thus we cannot reuse the effective potential results of ref.~\cite{Niemi:2021qvp}.

While there are only two vacuum topologies at two loops, there are a large number of fields in the xSM, so we automate the calculation using {\tt FeynRules}~\cite{Alloul:2013bka}, {\tt FeynArts}~\cite{Hahn:2000kx}, {\tt FeynCalc}~\cite{Shtabovenko:2016sxi} and {\tt Tarcer}~\cite{Mertig:1998vk}. The only required additional input is a two-loop master integral, the sunset diagram in 3d~\cite{Kajantie:1995dw}.

%%%%%%%%%%%%% Numerics %%%%%%%%%%%%%
\subsection{Numerics and tests} \label{appendix:numerics}

The perturbative approach outlined in Section~\ref{sec:factorisation} for the homogeneous equilibrium thermodynamics is relatively computationally cheap to evaluate. The extrema of the potential are solved at leading order within the EFT. This involves solving two coupled polynomial equations, and can be carried out analytically, yielding 9 solutions. To find the global minimum of the potential at a given temperature then simply requires evaluating the potential at the known analytical extrema.

If there is a change of phase between temperatures, we then compute the critical temperature and latent heat order-by-order in loops within the 3d EFT. Making use of the factorisation of scales, the analysis is carried out for fixed matching relations, either leading order or including the subleading $x^2$ corrections.

To solve equation~\eqref{eq:saddlepoint} for the critical bubble, we use {\tt CosmoTransitions}~\cite{Wainwright:2011kj}. This can fail for temperatures very close to $T_\text{c}$, so we first compute the surface tension at the critical temperature and then use a thin-wall approximation for temperatures very close to $T_\text{c}$, such that our thin-wall estimate gives $S_3[\varphi] > 2000$.

As evidence that our numerical implementation was doing what we expected, we tested and ensured the following:
\begin{enumerate}[(i)]
    \item Minimisation of the potential and perturbative expansion to machine precision.
    \item Agreement to machine precision with analytic leading-order results for the critical temperature and surface tension in the Yukawa model~\cite{Gould:2023jbz}.
    \item Agreement to machine precision for the one-loop \MSbar\ matching relations, the renormalisation group running, the dimensional reduction matching relations and the two-loop effective potential (with temporal gauge fields dropped) with Mathematica implementations for ref.~\cite{Niemi:2021qvp} shared by L.~Niemi and T.V.I.~Tenkanen.
    \item Approximate agreement with all the results of ref.~\cite{Gould:2021oba} at the two benchmark points studied there.
    \item Approximate agreement of the gravitational wave peak frequency and amplitude with {\tt PTPlot v1.1.0}~\cite{Caprini:2019egz}.
\end{enumerate}
Further details of our numerical implementation can be found in the source code~\cite{cosmoxsm}.

%%%%%%%%%%%% end %%%%%%%%%%%%

\bibliographystyle{JHEP}
\bibliography{refs.bib}

\providecommand{\href}[2]{#2}\begingroup\raggedright\begin{thebibliography}{100}

\bibitem{Croon:2020cgk}
D.~Croon, O.~Gould, P.~Schicho, T.V.I.~Tenkanen and G.~White, \emph{{Theoretical uncertainties for cosmological first-order phase transitions}}, \href{https://doi.org/10.1007/JHEP04(2021)055}{\emph{JHEP} {\bfseries 04} (2021) 055} [\href{https://arxiv.org/abs/2009.10080}{{\ttfamily 2009.10080}}].

\bibitem{Carena:2019une}
M.~Carena, Z.~Liu and Y.~Wang, \emph{{Electroweak phase transition with spontaneous Z$_{2}$-breaking}}, \href{https://doi.org/10.1007/JHEP08(2020)107}{\emph{JHEP} {\bfseries 08} (2020) 107} [\href{https://arxiv.org/abs/1911.10206}{{\ttfamily 1911.10206}}].

\bibitem{Gould:2021oba}
O.~Gould and T.V.I.~Tenkanen, \emph{{On the perturbative expansion at high temperature and implications for cosmological phase transitions}}, \href{https://doi.org/10.1007/JHEP06(2021)069}{\emph{JHEP} {\bfseries 06} (2021) 069} [\href{https://arxiv.org/abs/2104.04399}{{\ttfamily 2104.04399}}].

\bibitem{Athron:2022jyi}
P.~Athron, C.~Balazs, A.~Fowlie, L.~Morris, G.~White and Y.~Zhang, \emph{{How arbitrary are perturbative calculations of the electroweak phase transition?}}, \href{https://doi.org/10.1007/JHEP01(2023)050}{\emph{JHEP} {\bfseries 01} (2023) 050} [\href{https://arxiv.org/abs/2208.01319}{{\ttfamily 2208.01319}}].

\bibitem{Lewicki:2024xan}
M.~Lewicki, M.~Merchand, L.~Sagunski, P.~Schicho and D.~Schmitt, \emph{{Impact of theoretical uncertainties on model parameter reconstruction from GW signals sourced by cosmological phase transitions}},  \href{https://arxiv.org/abs/2403.03769}{{\ttfamily 2403.03769}}.

\bibitem{Ramsey-Musolf:2024ykk}
M.J.~Ramsey-Musolf, T.V.I.~Tenkanen and V.Q.~Tran, \emph{{Refining Gravitational Wave and Collider Physics Dialogue via Singlet Scalar Extension}},  \href{https://arxiv.org/abs/2409.17554}{{\ttfamily 2409.17554}}.

\bibitem{Kajantie:1995kf}
K.~Kajantie, M.~Laine, K.~Rummukainen and M.E.~Shaposhnikov, \emph{{The Electroweak phase transition: A Nonperturbative analysis}}, \href{https://doi.org/10.1016/0550-3213(96)00052-1}{\emph{Nucl. Phys. B} {\bfseries 466} (1996) 189} [\href{https://arxiv.org/abs/hep-lat/9510020}{{\ttfamily hep-lat/9510020}}].

\bibitem{Kajantie:1996mn}
K.~Kajantie, M.~Laine, K.~Rummukainen and M.E.~Shaposhnikov, \emph{{Is there a~ hot electroweak phase transition at $m_H \gtrsim m_W$?}}, \href{https://doi.org/10.1103/PhysRevLett.77.2887}{\emph{Phys. Rev. Lett.} {\bfseries 77} (1996) 2887} [\href{https://arxiv.org/abs/hep-ph/9605288}{{\ttfamily hep-ph/9605288}}].

\bibitem{Linde:1980ts}
A.D.~Linde, \emph{{Infrared Problem in Thermodynamics of the Yang-Mills Gas}}, \href{https://doi.org/10.1016/0370-2693(80)90769-8}{\emph{Phys. Lett. B} {\bfseries 96} (1980) 289}.

\bibitem{Kajantie:1995dw}
K.~Kajantie, M.~Laine, K.~Rummukainen and M.E.~Shaposhnikov, \emph{{Generic rules for high temperature dimensional reduction and their application to the standard model}}, \href{https://doi.org/10.1016/0550-3213(95)00549-8}{\emph{Nucl. Phys. B} {\bfseries 458} (1996) 90} [\href{https://arxiv.org/abs/hep-ph/9508379}{{\ttfamily hep-ph/9508379}}].

\bibitem{Ekstedt:2022zro}
A.~Ekstedt, O.~Gould and J.~L\"ofgren, \emph{{Radiative first-order phase transitions to next-to-next-to-leading order}}, \href{https://doi.org/10.1103/PhysRevD.106.036012}{\emph{Phys. Rev. D} {\bfseries 106} (2022) 036012} [\href{https://arxiv.org/abs/2205.07241}{{\ttfamily 2205.07241}}].

\bibitem{Ekstedt:2024etx}
A.~Ekstedt, P.~Schicho and T.V.I.~Tenkanen, \emph{{Cosmological phase transitions at three loops: the final verdict on perturbation theory}},  \href{https://arxiv.org/abs/2405.18349}{{\ttfamily 2405.18349}}.

\bibitem{Gould:2022ran}
O.~Gould, S.~G\"uyer and K.~Rummukainen, \emph{{First-order electroweak phase transitions: A nonperturbative update}}, \href{https://doi.org/10.1103/PhysRevD.106.114507}{\emph{Phys. Rev. D} {\bfseries 106} (2022) 114507} [\href{https://arxiv.org/abs/2205.07238}{{\ttfamily 2205.07238}}].

\bibitem{Gould:2023jbz}
O.~Gould and C.~Xie, \emph{{Higher orders for cosmological phase transitions: a global study in a Yukawa model}}, \href{https://doi.org/10.1007/JHEP12(2023)049}{\emph{JHEP} {\bfseries 12} (2023) 049} [\href{https://arxiv.org/abs/2310.02308}{{\ttfamily 2310.02308}}].

\bibitem{Niemi:2021qvp}
L.~Niemi, P.~Schicho and T.V.I.~Tenkanen, \emph{{Singlet-assisted electroweak phase transition at two loops}}, \href{https://doi.org/10.1103/PhysRevD.103.115035}{\emph{Phys. Rev. D} {\bfseries 103} (2021) 115035} [\href{https://arxiv.org/abs/2103.07467}{{\ttfamily 2103.07467}}].

\bibitem{Silveira:1985rk}
V.~Silveira and A.~Zee, \emph{{Scalar phantoms}}, \href{https://doi.org/10.1016/0370-2693(85)90624-0}{\emph{Phys. Lett. B} {\bfseries 161} (1985) 136}.

\bibitem{GAMBIT:2017gge}
{\scshape GAMBIT} collaboration, \emph{{Status of the scalar singlet dark matter model}}, \href{https://doi.org/10.1140/epjc/s10052-017-5113-1}{\emph{Eur. Phys. J. C} {\bfseries 77} (2017) 568} [\href{https://arxiv.org/abs/1705.07931}{{\ttfamily 1705.07931}}].

\bibitem{Cline:2012hg}
J.M.~Cline and K.~Kainulainen, \emph{{Electroweak baryogenesis and dark matter from a singlet Higgs}}, \href{https://doi.org/10.1088/1475-7516/2013/01/012}{\emph{JCAP} {\bfseries 01} (2013) 012} [\href{https://arxiv.org/abs/1210.4196}{{\ttfamily 1210.4196}}].

\bibitem{Cline:2021iff}
J.M.~Cline, A.~Friedlander, D.-M.~He, K.~Kainulainen, B.~Laurent and D.~Tucker-Smith, \emph{{Baryogenesis and gravity waves from a UV-completed electroweak phase transition}}, \href{https://doi.org/10.1103/PhysRevD.103.123529}{\emph{Phys. Rev. D} {\bfseries 103} (2021) 123529} [\href{https://arxiv.org/abs/2102.12490}{{\ttfamily 2102.12490}}].

\bibitem{Harigaya:2022ptp}
K.~Harigaya and I.R.~Wang, \emph{{First-Order Electroweak Phase Transition and Baryogenesis from a Naturally Light Singlet Scalar}},  \href{https://arxiv.org/abs/2207.02867}{{\ttfamily 2207.02867}}.

\bibitem{Ellis:2022lft}
J.~Ellis, M.~Lewicki, M.~Merchand, J.M.~No and M.~Zych, \emph{{The scalar singlet extension of the Standard Model: gravitational waves versus baryogenesis}}, \href{https://doi.org/10.1007/JHEP01(2023)093}{\emph{JHEP} {\bfseries 01} (2023) 093} [\href{https://arxiv.org/abs/2210.16305}{{\ttfamily 2210.16305}}].

\bibitem{Burrage:2018dvt}
C.~Burrage, E.J.~Copeland, P.~Millington and M.~Spannowsky, \emph{{Fifth forces, Higgs portals and broken scale invariance}}, \href{https://doi.org/10.1088/1475-7516/2018/11/036}{\emph{JCAP} {\bfseries 11} (2018) 036} [\href{https://arxiv.org/abs/1804.07180}{{\ttfamily 1804.07180}}].

\bibitem{Brax:2021rwk}
P.~Brax and C.~Burrage, \emph{{Screening the Higgs portal}}, \href{https://doi.org/10.1103/PhysRevD.104.015011}{\emph{Phys. Rev. D} {\bfseries 104} (2021) 015011} [\href{https://arxiv.org/abs/2101.10693}{{\ttfamily 2101.10693}}].

\bibitem{DOnofrio:2015gop}
M.~D'Onofrio and K.~Rummukainen, \emph{{Standard model cross-over on the lattice}}, \href{https://doi.org/10.1103/PhysRevD.93.025003}{\emph{Phys. Rev. D} {\bfseries 93} (2016) 025003} [\href{https://arxiv.org/abs/1508.07161}{{\ttfamily 1508.07161}}].

\bibitem{Arnold:1992rz}
P.B.~Arnold and O.~Espinosa, \emph{{The Effective potential and first order phase transitions: Beyond leading-order}}, \href{https://doi.org/10.1103/PhysRevD.47.3546}{\emph{Phys. Rev. D} {\bfseries 47} (1993) 3546} [\href{https://arxiv.org/abs/hep-ph/9212235}{{\ttfamily hep-ph/9212235}}].

\bibitem{Schicho:2021gca}
P.M.~Schicho, T.V.I.~Tenkanen and J.~\"Osterman, \emph{{Robust approach to thermal resummation: Standard Model meets a singlet}}, \href{https://doi.org/10.1007/JHEP06(2021)130}{\emph{JHEP} {\bfseries 06} (2021) 130} [\href{https://arxiv.org/abs/2102.11145}{{\ttfamily 2102.11145}}].

\bibitem{Niemi:2024vzw}
L.~Niemi and T.V.I.~Tenkanen, \emph{{Investigating two-loop effects for first-order electroweak phase transitions}},  \href{https://arxiv.org/abs/2408.15912}{{\ttfamily 2408.15912}}.

\bibitem{Niemi:2024axp}
L.~Niemi, M.J.~Ramsey-Musolf and G.~Xia, \emph{{Nonperturbative study of the electroweak phase transition in the real scalar singlet extended Standard Model}},  \href{https://arxiv.org/abs/2405.01191}{{\ttfamily 2405.01191}}.

\bibitem{Blaizot:2003iq}
J.P.~Blaizot, E.~Iancu and A.~Rebhan, \emph{{On the apparent convergence of perturbative QCD at high temperature}}, \href{https://doi.org/10.1103/PhysRevD.68.025011}{\emph{Phys. Rev. D} {\bfseries 68} (2003) 025011} [\href{https://arxiv.org/abs/hep-ph/0303045}{{\ttfamily hep-ph/0303045}}].

\bibitem{Laine:2006cp}
M.~Laine and Y.~Schroder, \emph{{Quark mass thresholds in QCD thermodynamics}}, \href{https://doi.org/10.1103/PhysRevD.73.085009}{\emph{Phys. Rev. D} {\bfseries 73} (2006) 085009} [\href{https://arxiv.org/abs/hep-ph/0603048}{{\ttfamily hep-ph/0603048}}].

\bibitem{Gould:2023ovu}
O.~Gould and T.V.I.~Tenkanen, \emph{{Perturbative effective field theory expansions for cosmological phase transitions}}, \href{https://doi.org/10.1007/JHEP01(2024)048}{\emph{JHEP} {\bfseries 01} (2024) 048} [\href{https://arxiv.org/abs/2309.01672}{{\ttfamily 2309.01672}}].

\bibitem{Braaten:1995cm}
E.~Braaten and A.~Nieto, \emph{{Effective field theory approach to high temperature thermodynamics}}, \href{https://doi.org/10.1103/PhysRevD.51.6990}{\emph{Phys. Rev. D} {\bfseries 51} (1995) 6990} [\href{https://arxiv.org/abs/hep-ph/9501375}{{\ttfamily hep-ph/9501375}}].

\bibitem{Bender:1999box}
C.M.~Bender and S.A.~Orszag, \emph{{Advanced Mathematical Methods for Scientists and Engineers I}}, Springer (1999), \href{https://doi.org/10.1007/978-1-4757-3069-2}{10.1007/978-1-4757-3069-2}.

\bibitem{Lofgren:2023sep}
J.~L\"ofgren, \emph{{Stop comparing resummation methods}}, \href{https://doi.org/10.1088/1361-6471/ad074b}{\emph{J. Phys. G} {\bfseries 50} (2023) 125008} [\href{https://arxiv.org/abs/2301.05197}{{\ttfamily 2301.05197}}].

\bibitem{Camargo-Molina:2024sde}
E.~Camargo-Molina, R.~Enberg and J.~L\"ofgren, \emph{{A Catalog of First-Order Electroweak Phase Transitions in the Standard Model Effective Field Theory}},  \href{https://arxiv.org/abs/2410.23210}{{\ttfamily 2410.23210}}.

\bibitem{Ekstedt:2020abj}
A.~Ekstedt and J.~L\"ofgren, \emph{{A Critical Look at the Electroweak Phase Transition}}, \href{https://doi.org/10.1007/JHEP12(2020)136}{\emph{JHEP} {\bfseries 12} (2020) 136} [\href{https://arxiv.org/abs/2006.12614}{{\ttfamily 2006.12614}}].

\bibitem{Camargo-Molina:2021zgz}
J.E.~Camargo-Molina, R.~Enberg and J.~L\"ofgren, \emph{{A new perspective on the electroweak phase transition in the Standard Model Effective Field Theory}}, \href{https://doi.org/10.1007/JHEP10(2021)127}{\emph{JHEP} {\bfseries 10} (2021) 127} [\href{https://arxiv.org/abs/2103.14022}{{\ttfamily 2103.14022}}].

\bibitem{Kierkla:2023von}
M.~Kierkla, B.~Swiezewska, T.V.I.~Tenkanen and J.~van~de Vis, \emph{{Gravitational waves from supercooled phase transitions: dimensional transmutation meets dimensional reduction}}, \href{https://doi.org/10.1007/JHEP02(2024)234}{\emph{JHEP} {\bfseries 02} (2024) 234} [\href{https://arxiv.org/abs/2312.12413}{{\ttfamily 2312.12413}}].

\bibitem{Brauner:2016fla}
T.~Brauner, T.V.I.~Tenkanen, A.~Tranberg, A.~Vuorinen and D.J.~Weir, \emph{{Dimensional reduction of the Standard Model coupled to a new singlet scalar field}}, \href{https://doi.org/10.1007/JHEP03(2017)007}{\emph{JHEP} {\bfseries 03} (2017) 007} [\href{https://arxiv.org/abs/1609.06230}{{\ttfamily 1609.06230}}].

\bibitem{Chala:2024xll}
M.~Chala, J.C.~Criado, L.~Gil and J.L.~Miras, \emph{{Higher-order corrections to phase-transition parameters in dimensional reduction}},  \href{https://arxiv.org/abs/2406.02667}{{\ttfamily 2406.02667}}.

\bibitem{Caprini:2019egz}
C.~Caprini et~al., \emph{{Detecting gravitational waves from cosmological phase transitions with LISA: an update}}, \href{https://doi.org/10.1088/1475-7516/2020/03/024}{\emph{JCAP} {\bfseries 03} (2020) 024} [\href{https://arxiv.org/abs/1910.13125}{{\ttfamily 1910.13125}}].

\bibitem{Caprini:2024hue}
{\scshape LISA Cosmology Working Group} collaboration, \emph{{Gravitational waves from first-order phase transitions in LISA: reconstruction pipeline and physics interpretation}}, \href{https://doi.org/10.1088/1475-7516/2024/10/020}{\emph{JCAP} {\bfseries 10} (2024) 020} [\href{https://arxiv.org/abs/2403.03723}{{\ttfamily 2403.03723}}].

\bibitem{Wang:2024slx}
X.~Wang, C.~Tian and C.~Bal\'azs, \emph{{Self-consistent prediction of gravitational waves from cosmological phase transitions}},  \href{https://arxiv.org/abs/2409.06599}{{\ttfamily 2409.06599}}.

\bibitem{Tian:2024ysd}
C.~Tian, X.~Wang and C.~Bal\'azs, \emph{{Gravitational waves from cosmological first-order phase transitions with precise hydrodynamics}},  \href{https://arxiv.org/abs/2409.14505}{{\ttfamily 2409.14505}}.

\bibitem{Hindmarsh:2017gnf}
M.~Hindmarsh, S.J.~Huber, K.~Rummukainen and D.J.~Weir, \emph{{Shape of the acoustic gravitational wave power spectrum from a first order phase transition}}, \href{https://doi.org/10.1103/PhysRevD.96.103520}{\emph{Phys. Rev. D} {\bfseries 96} (2017) 103520} [\href{https://arxiv.org/abs/1704.05871}{{\ttfamily 1704.05871}}].

\bibitem{Jinno:2022mie}
R.~Jinno, T.~Konstandin, H.~Rubira and I.~Stomberg, \emph{{Higgsless simulations of cosmological phase transitions and gravitational waves}}, \href{https://doi.org/10.1088/1475-7516/2023/02/011}{\emph{JCAP} {\bfseries 02} (2023) 011} [\href{https://arxiv.org/abs/2209.04369}{{\ttfamily 2209.04369}}].

\bibitem{Giese:2020znk}
F.~Giese, T.~Konstandin, K.~Schmitz and J.~van~de Vis, \emph{{Model-independent energy budget for LISA}}, \href{https://doi.org/10.1088/1475-7516/2021/01/072}{\emph{JCAP} {\bfseries 01} (2021) 072} [\href{https://arxiv.org/abs/2010.09744}{{\ttfamily 2010.09744}}].

\bibitem{Ai:2023see}
W.-Y.~Ai, B.~Laurent and J.~van~de Vis, \emph{{Model-independent bubble wall velocities in local thermal equilibrium}}, \href{https://doi.org/10.1088/1475-7516/2023/07/002}{\emph{JCAP} {\bfseries 07} (2023) 002} [\href{https://arxiv.org/abs/2303.10171}{{\ttfamily 2303.10171}}].

\bibitem{Guo:2021qcq}
H.-K.~Guo, K.~Sinha, D.~Vagie and G.~White, \emph{{The benefits of diligence: how precise are predicted gravitational wave spectra in models with phase transitions?}}, \href{https://doi.org/10.1007/JHEP06(2021)164}{\emph{JHEP} {\bfseries 06} (2021) 164} [\href{https://arxiv.org/abs/2103.06933}{{\ttfamily 2103.06933}}].

\bibitem{Athron:2023rfq}
P.~Athron, L.~Morris and Z.~Xu, \emph{{How robust are gravitational wave predictions from cosmological phase transitions?}}, \href{https://doi.org/10.1088/1475-7516/2024/05/075}{\emph{JCAP} {\bfseries 05} (2024) 075} [\href{https://arxiv.org/abs/2309.05474}{{\ttfamily 2309.05474}}].

\bibitem{Nielsen:1975fs}
N.K.~Nielsen, \emph{{On the Gauge Dependence of Spontaneous Symmetry Breaking in Gauge Theories}}, \href{https://doi.org/10.1016/0550-3213(75)90301-6}{\emph{Nucl. Phys. B} {\bfseries 101} (1975) 173}.

\bibitem{Fukuda:1975di}
R.~Fukuda and T.~Kugo, \emph{{Gauge Invariance in the Effective Action and Potential}}, \href{https://doi.org/10.1103/PhysRevD.13.3469}{\emph{Phys. Rev. D} {\bfseries 13} (1976) 3469}.

\bibitem{Laine:1994zq}
M.~Laine, \emph{{Gauge dependence of the high temperature two loop effective potential for the Higgs field}}, \href{https://doi.org/10.1103/PhysRevD.51.4525}{\emph{Phys. Rev. D} {\bfseries 51} (1995) 4525} [\href{https://arxiv.org/abs/hep-ph/9411252}{{\ttfamily hep-ph/9411252}}].

\bibitem{Enqvist:1991xw}
K.~Enqvist, J.~Ignatius, K.~Kajantie and K.~Rummukainen, \emph{{Nucleation and bubble growth in a first order cosmological electroweak phase transition}}, \href{https://doi.org/10.1103/PhysRevD.45.3415}{\emph{Phys. Rev. D} {\bfseries 45} (1992) 3415}.

\bibitem{Tenkanen:2022tly}
T.V.I.~Tenkanen and J.~van~de Vis, \emph{{Speed of sound in cosmological phase transitions and effect on gravitational waves}}, \href{https://doi.org/10.1007/JHEP08(2022)302}{\emph{JHEP} {\bfseries 08} (2022) 302} [\href{https://arxiv.org/abs/2206.01130}{{\ttfamily 2206.01130}}].

\bibitem{Laurent:2022jrs}
B.~Laurent and J.M.~Cline, \emph{{First principles determination of bubble wall velocity}}, \href{https://doi.org/10.1103/PhysRevD.106.023501}{\emph{Phys. Rev. D} {\bfseries 106} (2022) 023501} [\href{https://arxiv.org/abs/2204.13120}{{\ttfamily 2204.13120}}].

\bibitem{Hindmarsh:2015qta}
M.~Hindmarsh, S.J.~Huber, K.~Rummukainen and D.J.~Weir, \emph{{Numerical simulations of acoustically generated gravitational waves at a first order phase transition}}, \href{https://doi.org/10.1103/PhysRevD.92.123009}{\emph{Phys. Rev. D} {\bfseries 92} (2015) 123009} [\href{https://arxiv.org/abs/1504.03291}{{\ttfamily 1504.03291}}].

\bibitem{Guo:2020grp}
H.-K.~Guo, K.~Sinha, D.~Vagie and G.~White, \emph{{Phase Transitions in an Expanding Universe: Stochastic Gravitational Waves in Standard and Non-Standard Histories}}, \href{https://doi.org/10.1088/1475-7516/2021/01/001}{\emph{JCAP} {\bfseries 01} (2021) 001} [\href{https://arxiv.org/abs/2007.08537}{{\ttfamily 2007.08537}}].

\bibitem{Gould:2021dzl}
O.~Gould, \emph{{Real scalar phase transitions: a nonperturbative analysis}}, \href{https://doi.org/10.1007/JHEP04(2021)057}{\emph{JHEP} {\bfseries 04} (2021) 057} [\href{https://arxiv.org/abs/2101.05528}{{\ttfamily 2101.05528}}].

\bibitem{Hirvonen:2021zej}
J.~Hirvonen, J.~L\"ofgren, M.J.~Ramsey-Musolf, P.~Schicho and T.V.I.~Tenkanen, \emph{{Computing the gauge-invariant bubble nucleation rate in finite temperature effective field theory}}, \href{https://doi.org/10.1007/JHEP07(2022)135}{\emph{JHEP} {\bfseries 07} (2022) 135} [\href{https://arxiv.org/abs/2112.08912}{{\ttfamily 2112.08912}}].

\bibitem{Kajantie:2002wa}
K.~Kajantie, M.~Laine, K.~Rummukainen and Y.~Schroder, \emph{{The Pressure of hot QCD up to g6 ln(1/g)}}, \href{https://doi.org/10.1103/PhysRevD.67.105008}{\emph{Phys. Rev. D} {\bfseries 67} (2003) 105008} [\href{https://arxiv.org/abs/hep-ph/0211321}{{\ttfamily hep-ph/0211321}}].

\bibitem{Weinberg:1980wa}
S.~Weinberg, \emph{{Effective Gauge Theories}}, \href{https://doi.org/10.1016/0370-2693(80)90660-7}{\emph{Phys. Lett. B} {\bfseries 91} (1980) 51}.

\bibitem{Gould:2021ccf}
O.~Gould and J.~Hirvonen, \emph{{Effective field theory approach to thermal bubble nucleation}}, \href{https://doi.org/10.1103/PhysRevD.104.096015}{\emph{Phys. Rev. D} {\bfseries 104} (2021) 096015} [\href{https://arxiv.org/abs/2108.04377}{{\ttfamily 2108.04377}}].

\bibitem{Langer:1969bc}
J.S.~Langer, \emph{{Statistical theory of the decay of metastable states}}, \href{https://doi.org/10.1016/0003-4916(69)90153-5}{\emph{Annals Phys.} {\bfseries 54} (1969) 258}.

\bibitem{Affleck:1980ac}
I.~Affleck, \emph{{Quantum Statistical Metastability}}, \href{https://doi.org/10.1103/PhysRevLett.46.388}{\emph{Phys. Rev. Lett.} {\bfseries 46} (1981) 388}.

\bibitem{Linde:1981zj}
A.D.~Linde, \emph{{Decay of the False Vacuum at Finite Temperature}}, \href{https://doi.org/10.1016/0550-3213(83)90072-X}{\emph{Nucl. Phys. B} {\bfseries 216} (1983) 421}.

\bibitem{Langer:1974cpa}
J.S.~Langer, \emph{{Metastable states}}, \href{https://doi.org/10.1016/0031-8914(74)90226-2}{\emph{Physica} {\bfseries 73} (1974) 61}.

\bibitem{Arnold:1987mh}
P.B.~Arnold and L.D.~McLerran, \emph{{Sphalerons, Small Fluctuations and Baryon Number Violation in Electroweak Theory}}, \href{https://doi.org/10.1103/PhysRevD.36.581}{\emph{Phys. Rev. D} {\bfseries 36} (1987) 581}.

\bibitem{Hanggi:1990zz}
P.~Hanggi, P.~Talkner and M.~Borkovec, \emph{{Reaction-Rate Theory: Fifty Years After Kramers}}, \href{https://doi.org/10.1103/RevModPhys.62.251}{\emph{Rev. Mod. Phys.} {\bfseries 62} (1990) 251}.

\bibitem{Bodeker:1998hm}
D.~Bodeker, \emph{{On the effective dynamics of soft nonAbelian gauge fields at finite temperature}}, \href{https://doi.org/10.1016/S0370-2693(98)00279-2}{\emph{Phys. Lett. B} {\bfseries 426} (1998) 351} [\href{https://arxiv.org/abs/hep-ph/9801430}{{\ttfamily hep-ph/9801430}}].

\bibitem{Ekstedt:2022tqk}
A.~Ekstedt, \emph{{Bubble nucleation to all orders}}, \href{https://doi.org/10.1007/JHEP08(2022)115}{\emph{JHEP} {\bfseries 08} (2022) 115} [\href{https://arxiv.org/abs/2201.07331}{{\ttfamily 2201.07331}}].

\bibitem{Hirvonen:2024rfg}
J.~Hirvonen, \emph{{Nucleation Rate in a High-Temperature Quantum Field Theory with Hard Particles}},  \href{https://arxiv.org/abs/2403.07987}{{\ttfamily 2403.07987}}.

\bibitem{Pirvu:2024ova}
D.~P\^\i{}rvu, A.~Shkerin and S.~Sibiryakov, \emph{{Thermal False Vacuum Decay Is Not What It Seems}},  \href{https://arxiv.org/abs/2407.06263}{{\ttfamily 2407.06263}}.

\bibitem{Pirvu:2024nbe}
D.~P\^\i{}rvu, A.~Shkerin and S.~Sibiryakov, \emph{{Thermal false vacuum decay in (1+1)-dimensions: Evidence for non-equilibrium dynamics}},  \href{https://arxiv.org/abs/2408.06411}{{\ttfamily 2408.06411}}.

\bibitem{Gould:2024chm}
O.~Gould, A.~Kormu and D.J.~Weir, \emph{{A nonperturbative test of nucleation calculations for strong phase transitions}},  \href{https://arxiv.org/abs/2404.01876}{{\ttfamily 2404.01876}}.

\bibitem{Bezuglov:2018qpq}
M.A.~Bezuglov and A.I.~Onishchenko, \emph{{Two-loop corrections to false vacuum decay in scalar field theory}}, \href{https://doi.org/10.1016/j.physletb.2018.11.005}{\emph{Phys. Lett. B} {\bfseries 788} (2019) 122} [\href{https://arxiv.org/abs/1805.06482}{{\ttfamily 1805.06482}}].

\bibitem{Konstandin:2010dm}
T.~Konstandin and J.M.~No, \emph{{Hydrodynamic obstruction to bubble expansion}}, \href{https://doi.org/10.1088/1475-7516/2011/02/008}{\emph{JCAP} {\bfseries 02} (2011) 008} [\href{https://arxiv.org/abs/1011.3735}{{\ttfamily 1011.3735}}].

\bibitem{BarrosoMancha:2020fay}
M.~Barroso~Mancha, T.~Prokopec and B.~Swiezewska, \emph{{Field-theoretic derivation of bubble-wall force}}, \href{https://doi.org/10.1007/JHEP01(2021)070}{\emph{JHEP} {\bfseries 01} (2021) 070} [\href{https://arxiv.org/abs/2005.10875}{{\ttfamily 2005.10875}}].

\bibitem{Balaji:2020yrx}
S.~Balaji, M.~Spannowsky and C.~Tamarit, \emph{{Cosmological bubble friction in local equilibrium}}, \href{https://doi.org/10.1088/1475-7516/2021/03/051}{\emph{JCAP} {\bfseries 03} (2021) 051} [\href{https://arxiv.org/abs/2010.08013}{{\ttfamily 2010.08013}}].

\bibitem{Ai:2021kak}
W.-Y.~Ai, B.~Garbrecht and C.~Tamarit, \emph{{Bubble wall velocities in local equilibrium}}, \href{https://doi.org/10.1088/1475-7516/2022/03/015}{\emph{JCAP} {\bfseries 03} (2022) 015} [\href{https://arxiv.org/abs/2109.13710}{{\ttfamily 2109.13710}}].

\bibitem{chapman1952mathematical}
S.~Chapman, \emph{The mathematical theory of non-uniform gases; an account of the}, Cambridge University Press, Cambridge, UK (1952).

\bibitem{Dashko:2024spj}
A.~Dashko and A.~Ekstedt, \emph{{Bubble-wall speed with loop corrections}},  \href{https://arxiv.org/abs/2411.05075}{{\ttfamily 2411.05075}}.

\bibitem{DeCurtis:2024hvh}
S.~De~Curtis, L.~Delle~Rose, A.~Guiggiani, A.~Gil~Muyor and G.~Panico, \emph{{Non-linearities in cosmological bubble wall dynamics}}, \href{https://doi.org/10.1007/JHEP05(2024)009}{\emph{JHEP} {\bfseries 05} (2024) 009} [\href{https://arxiv.org/abs/2401.13522}{{\ttfamily 2401.13522}}].

\bibitem{Kainulainen:2019kyp}
K.~Kainulainen, V.~Keus, L.~Niemi, K.~Rummukainen, T.V.I.~Tenkanen and V.~Vaskonen, \emph{{On the validity of perturbative studies of the electroweak phase transition in the Two Higgs Doublet model}}, \href{https://doi.org/10.1007/JHEP06(2019)075}{\emph{JHEP} {\bfseries 06} (2019) 075} [\href{https://arxiv.org/abs/1904.01329}{{\ttfamily 1904.01329}}].

\bibitem{Espinosa:2011ax}
J.R.~Espinosa, T.~Konstandin and F.~Riva, \emph{{Strong Electroweak Phase Transitions in the Standard Model with a Singlet}}, \href{https://doi.org/10.1016/j.nuclphysb.2011.09.010}{\emph{Nucl. Phys. B} {\bfseries 854} (2012) 592} [\href{https://arxiv.org/abs/1107.5441}{{\ttfamily 1107.5441}}].

\bibitem{Blasi:2022woz}
S.~Blasi and A.~Mariotti, \emph{{Domain Walls Seeding the Electroweak Phase Transition}}, \href{https://doi.org/10.1103/PhysRevLett.129.261303}{\emph{Phys. Rev. Lett.} {\bfseries 129} (2022) 261303} [\href{https://arxiv.org/abs/2203.16450}{{\ttfamily 2203.16450}}].

\bibitem{Blasi:2023rqi}
S.~Blasi, R.~Jinno, T.~Konstandin, H.~Rubira and I.~Stomberg, \emph{{Gravitational waves from defect-driven phase transitions: domain walls}}, \href{https://doi.org/10.1088/1475-7516/2023/10/051}{\emph{JCAP} {\bfseries 10} (2023) 051} [\href{https://arxiv.org/abs/2302.06952}{{\ttfamily 2302.06952}}].

\bibitem{Agrawal:2023cgp}
P.~Agrawal, S.~Blasi, A.~Mariotti and M.~Nee, \emph{{Electroweak phase transition with a double well done doubly well}}, \href{https://doi.org/10.1007/JHEP06(2024)089}{\emph{JHEP} {\bfseries 06} (2024) 089} [\href{https://arxiv.org/abs/2312.06749}{{\ttfamily 2312.06749}}].

\bibitem{AbdusSalam:2020rdj}
S.S.~AbdusSalam et~al., \emph{{Simple and statistically sound recommendations for analysing physical theories}}, \href{https://doi.org/10.1088/1361-6633/ac60ac}{\emph{Rept. Prog. Phys.} {\bfseries 85} (2022) 052201} [\href{https://arxiv.org/abs/2012.09874}{{\ttfamily 2012.09874}}].

\bibitem{Ramsey-Musolf:2019lsf}
M.J.~Ramsey-Musolf, \emph{{The electroweak phase transition: a collider target}}, \href{https://doi.org/10.1007/JHEP09(2020)179}{\emph{JHEP} {\bfseries 09} (2020) 179} [\href{https://arxiv.org/abs/1912.07189}{{\ttfamily 1912.07189}}].

\bibitem{Alves:2018jsw}
A.~Alves, T.~Ghosh, H.-K.~Guo, K.~Sinha and D.~Vagie, \emph{{Collider and Gravitational Wave Complementarity in Exploring the Singlet Extension of the Standard Model}}, \href{https://doi.org/10.1007/JHEP04(2019)052}{\emph{JHEP} {\bfseries 04} (2019) 052} [\href{https://arxiv.org/abs/1812.09333}{{\ttfamily 1812.09333}}].

\bibitem{Aboudonia:2024frg}
M.~Aboudonia, C.~Balazs, A.~Papaefstathiou and G.~White, \emph{{Investigating the Electroweak Phase Transition with a Real Scalar Singlet at a Muon Collider}},  \href{https://arxiv.org/abs/2410.22700}{{\ttfamily 2410.22700}}.

\bibitem{gould_2024_14031507}
O.~Gould and P.~Saffin, ``{Perturbative gravitational wave predictions for the real scalar extended Standard Model, dataset}.'' \url{https://doi.org/10.5281/zenodo.14031507}, 11, 2024.
\newblock 10.5281/zenodo.14031507.

\bibitem{Gould:2019qek}
O.~Gould, J.~Kozaczuk, L.~Niemi, M.J.~Ramsey-Musolf, T.V.I.~Tenkanen and D.J.~Weir, \emph{{Nonperturbative analysis of the gravitational waves from a first-order electroweak phase transition}}, \href{https://doi.org/10.1103/PhysRevD.100.115024}{\emph{Phys. Rev. D} {\bfseries 100} (2019) 115024} [\href{https://arxiv.org/abs/1903.11604}{{\ttfamily 1903.11604}}].

\bibitem{Papaefstathiou:2020iag}
A.~Papaefstathiou and G.~White, \emph{{The electro-weak phase transition at colliders: confronting theoretical uncertainties and complementary channels}}, \href{https://doi.org/10.1007/JHEP05(2021)099}{\emph{JHEP} {\bfseries 05} (2021) 099} [\href{https://arxiv.org/abs/2010.00597}{{\ttfamily 2010.00597}}].

\bibitem{Schmitz:2020syl}
K.~Schmitz, \emph{{New Sensitivity Curves for Gravitational-Wave Signals from Cosmological Phase Transitions}}, \href{https://doi.org/10.1007/JHEP01(2021)097}{\emph{JHEP} {\bfseries 01} (2021) 097} [\href{https://arxiv.org/abs/2002.04615}{{\ttfamily 2002.04615}}].

\bibitem{LISA:2017pwj}
{\scshape LISA} collaboration, \emph{{Laser Interferometer Space Antenna}},  \href{https://arxiv.org/abs/1702.00786}{{\ttfamily 1702.00786}}.

\bibitem{Kawamura:2011zz}
S.~Kawamura et~al., \emph{{The Japanese space gravitational wave antenna: DECIGO}}, \href{https://doi.org/10.1088/0264-9381/28/9/094011}{\emph{Class. Quant. Grav.} {\bfseries 28} (2011) 094011}.

\bibitem{Eichhorn:2020upj}
A.~Eichhorn, J.~Lumma, J.M.~Pawlowski, M.~Reichert and M.~Yamada, \emph{{Universal gravitational-wave signatures from heavy new physics in the electroweak sector}}, \href{https://doi.org/10.1088/1475-7516/2021/05/006}{\emph{JCAP} {\bfseries 05} (2021) 006} [\href{https://arxiv.org/abs/2010.00017}{{\ttfamily 2010.00017}}].

\bibitem{Auclair:2022jod}
P.~Auclair, C.~Caprini, D.~Cutting, M.~Hindmarsh, K.~Rummukainen, D.A.~Steer et~al., \emph{{Generation of grKierkla:2023vonavitational waves from freely decaying turbulence}}, \href{https://doi.org/10.1088/1475-7516/2022/09/029}{\emph{JCAP} {\bfseries 09} (2022) 029} [\href{https://arxiv.org/abs/2205.02588}{{\ttfamily 2205.02588}}].

\bibitem{Dahl:2024eup}
J.~Dahl, M.~Hindmarsh, K.~Rummukainen and D.~Weir, \emph{{Primordial acoustic turbulence: three-dimensional simulations and gravitational wave predictions}},  \href{https://arxiv.org/abs/2407.05826}{{\ttfamily 2407.05826}}.

\bibitem{Caprini:2024gyk}
C.~Caprini, R.~Jinno, T.~Konstandin, A.~Roper~Pol, H.~Rubira and I.~Stomberg, \emph{{Gravitational waves from decaying sources in strong phase transitions}},  \href{https://arxiv.org/abs/2409.03651}{{\ttfamily 2409.03651}}.

\bibitem{Hindmarsh:2024ttn}
M.~Hindmarsh, D.C.~Hooper, T.~Minkkinen and D.J.~Weir, \emph{{Recovering a phase transition signal in simulated LISA data with a modulated galactic foreground}},  \href{https://arxiv.org/abs/2406.04894}{{\ttfamily 2406.04894}}.

\bibitem{cosmoxsm}
O.~Gould and P.~Saffin, ``cosmoxsm.'' \url{https://bitbucket.org/og113/cosmoxsm/}, 11, 2024.

\bibitem{Alloul:2013bka}
A.~Alloul, N.D.~Christensen, C.~Degrande, C.~Duhr and B.~Fuks, \emph{{FeynRules 2.0 - A complete toolbox for tree-level phenomenology}}, \href{https://doi.org/10.1016/j.cpc.2014.04.012}{\emph{Comput. Phys. Commun.} {\bfseries 185} (2014) 2250} [\href{https://arxiv.org/abs/1310.1921}{{\ttfamily 1310.1921}}].

\bibitem{Hahn:2000kx}
T.~Hahn, \emph{{Generating Feynman diagrams and amplitudes with FeynArts 3}}, \href{https://doi.org/10.1016/S0010-4655(01)00290-9}{\emph{Comput. Phys. Commun.} {\bfseries 140} (2001) 418} [\href{https://arxiv.org/abs/hep-ph/0012260}{{\ttfamily hep-ph/0012260}}].

\bibitem{Shtabovenko:2016sxi}
V.~Shtabovenko, R.~Mertig and F.~Orellana, \emph{{New Developments in FeynCalc 9.0}}, \href{https://doi.org/10.1016/j.cpc.2016.06.008}{\emph{Comput. Phys. Commun.} {\bfseries 207} (2016) 432} [\href{https://arxiv.org/abs/1601.01167}{{\ttfamily 1601.01167}}].

\bibitem{Mertig:1998vk}
R.~Mertig and R.~Scharf, \emph{{TARCER: A Mathematica program for the reduction of two loop propagator integrals}}, \href{https://doi.org/10.1016/S0010-4655(98)00042-3}{\emph{Comput. Phys. Commun.} {\bfseries 111} (1998) 265} [\href{https://arxiv.org/abs/hep-ph/9801383}{{\ttfamily hep-ph/9801383}}].

\bibitem{Wainwright:2011kj}
C.L.~Wainwright, \emph{{CosmoTransitions: Computing Cosmological Phase Transition Temperatures and Bubble Profiles with Multiple Fields}}, \href{https://doi.org/10.1016/j.cpc.2012.04.004}{\emph{Comput. Phys. Commun.} {\bfseries 183} (2012) 2006} [\href{https://arxiv.org/abs/1109.4189}{{\ttfamily 1109.4189}}].

\end{thebibliography}\endgroup
\end{document}